\documentclass{aa}  
\usepackage{scrextend}
\usepackage{hyperref}
\usepackage{pdflscape}
\usepackage{xcolor}
\usepackage{lscape}
\usepackage{graphicx}
\usepackage{lscape}
\usepackage{placeins}
\usepackage{txfonts}
\newcommand{\kms}{km~s$^{-1}$}
\usepackage{txfonts}
\usepackage{siunitx}
\usepackage{enumitem}
\usepackage{wrapfig}
\usepackage{subfig}
\usepackage{multirow}
\usepackage{bm}

\usepackage{cleveref}
\usepackage[flushleft]{threeparttable}

\newcommand{\Msun}{M$_{\odot}$}
\newcommand{\Rsun}{R$_{\odot}$}
\newcommand{\Lsun}{L$_{\odot}$}
\def\Hii{H\,{\sc ii} }

\newcolumntype{R}{@{\extracolsep{3cm}}r@{\extracolsep{0pt}}}%

\begin{document}

   \title{On the origin of close massive binaries in the M17 star-forming region}

   \author{E. Bordier\inst{1,2}, A. J. Frost\inst{1}, H. Sana\inst{1}, M. Reggiani\inst{1}, A. Mérand\inst{3}, A. Rainot\inst{1}, M.C. Ramírez-Tannus\inst{4}, W.J. de Wit\inst{2}}

   \institute{\inst{1}Institute of Astronomy, KU Leuven, Celestijnenlaan 200D, B-3001 Leuven, Belgium \\
              \email{emma.bordier@kuleuven.be} \\
              \inst{2} European Southern Observatory (ESO), Alonso de Cordova 3107, Vitacura, Santiago, Chile \\
              \inst{3} European Southern Observatory (ESO), Karl-Schwarzschild-Str. 2, Garching, Germany \\
              \inst{4}Max Planck Institute for Astronomy, Königstuhl 17, D-69117 Heidelberg, Germany \\}

   \date{Received Month xx, xxxx; accepted Month xx, xxxx}

   \titlerunning{On the origin of close massive binaries in the M17 star-forming region}
   \authorrunning{E.~Bordier et al.}

  \abstract
   {Spectroscopic multiplicity surveys of O stars in young clusters and OB associations have revealed that a large portion ($\sim$70~\%) of these massive stars ($M_\mathrm{i}>15$~\Msun) belong to close and short-period binaries (physical separation d $<$ few au). Follow-up VLT(I) high-angular resolution observations led to the detection of wider companions (up to d$\sim$500~au), increasing the average companion fraction to $>$2. Despite the recent and significant progress, the formation mechanisms leading to such close massive multiple systems remain to be elucidated. As a result, young massive close binaries (or higher-order multiple systems) are unique laboratories to figure out the pairing mechanism of high-mass stars. }
   {We present the first VLTI/GRAVITY observations of six young O stars in the M17 star-forming region ($\lesssim$1~Myr) and two additional foreground stars. VLTI/GRAVITY provides the $K$-band high-angular resolution observations needed to explore the close environment of young O-type stars and, as such, offers an excellent opportunity to characterise the multiplicity properties of the immediate outcome of the massive star formation process.}
   {From the interferometric model fitting of visibility amplitudes and closure phases, we search for companions and measure their positions and flux ratios. Combining the resulting magnitude difference with atmosphere models and evolutionary tracks, we further constrain the masses of the individual components.}
   {All of the six high-mass stars are in multiple systems, leading to a multiplicity fraction (MF) of 100\%, yielding a 68\% confidence interval of 94$-$100\%. We detect a total number of 9 companions with separations up to 120~au. Including previously identified spectroscopic companions, the companion fraction of the young O-stars in our sample reaches $2.3\pm0.6$. The derived masses span a wide range from 2.5 to 50~\Msun, with a great tendency towards high-mass companions. However, we do not find a significant correlation between the mass of the companions and their separation.}
    {While based on a modest sample, our results clearly indicate that the origin of the high degree of multiplicity is rooted in their star formation mechanism. No clear evidence for one of the competing concepts of massive star formation (core accretion or competitive accretion) could be found. However, given that we find all of the companions within $\sim$120~au, our results are compatible with migration as a scenario for the formation of close massive binaries.}

   \keywords{ Stars: massive -- Stars: early-type -- (Stars:) binaries (including multiple): close -- Stars: formation -- Techniques: interferometric}

   \maketitle

\section{Introduction}

From their births to their explosive deaths, massive stars (those with initial masses higher than 8\Msun) are known to play a key role in our Universe. As the main producers of alpha-elements and through their strong stellar winds and dramatic end-of-life explosions, they heat and enrich the interstellar medium, increasing the metallicity of the primordial gas and driving the chemical evolution of their host galaxy \citep{Bromm+2009}. 

Massive star evolution strongly depends on their multiplicity properties at birth \citep{Norbert2012}. Observationally, it is established that at least 90\% of massive stars are found within a binary or higher-order multiple system \citep{DeWit+2005,Sana+2014}. Most of them have at least one companion with an orbital period of the order of two months or shorter, that is close enough for the two stars to interact during their lifetime \citep{Sana+2012, Sana+2013}. These are efficiently detected by spectroscopy through the periodic Doppler shifts of their spectral lines. Such interactions are of paramount importance as they dramatically impact the evolution of the stars involved. 

Thus far, multiplicity surveys were dedicated to fully formed stars drawn from massive star populations with estimated age of typically 2-8~Myr. These do not necessarily reflect the primordial binary properties. To date, two main surveys targeting the multiplicity of pre-main sequence (PMS) stars have been conducted. \citet{Pomohaci+2019} surveyed 32 massive young stellar objects (MYSOs) with NACO/VLT, probing for wide binary companions ($>$600~au). They reported a 100\% fraction of MYSOs belonging to multiple systems, with mass ratios exceeding 0.5. They also claim that most of the MYSOs might be found in higher order multiple systems. More recently, \citet{Koumpia+2021} reported a binary fraction of 17-25\% at 2-300~au scales, among a population of six MYSOs observed with VLTI (GRAVITY and AMBER). Single objects studies also contribute to this growing number of exciting results: \citet{Kraus+2017} outlined the discovery of a high-massive protobinary system (20+18\Msun, $d\sim170$~au), each surrounded by a circumstellar accretion disk and a circumbinary disk. \citet{Zhang+2019} recently published the results of a similar object: a massive protobinary with an apparent separation of $180\pm11$~au and about 18~\Msun\ residing at the centre of the IRAS07299-1651 star-forming region is fed by two circumstellar disks around each protostar. Such results raise the question of the high-mass star formation and multiplicity properties at birth.

Unlike the case for low and intermediate mass stars, a definitive observational sequence for massive star formation is yet to be obtained. From the theoretical point of view, several formation scenarios have been proposed such as formation through stellar collisions and merging \citep{Bonnell+1998}, competitive accretion \citep{Bonnell+2001,Bonnell+2006}, fragmentation-induced starvation \citep{Peters+2010}, and monolithic collapse \citep{McKee+2003,Krumholz+2009}. Except for the merger process, the scenarios share the need for dense and massive accretion disks to overcome the radiation pressure barrier. Such disks are prone to gravitational instabilities \citep{Kratter+2010}, predicting companions at much larger separations than typically observed among massive spectroscopic binaries. Circumstellar disk sizes are of great interest to spot the formation of potential companions in the neighbourhood of the central star. Over the past decade an increasing number of disks have been detected around high-mass protostars, for which an estimation of the inner (from 10s to 100s~au, \citealt{Kraus+2010,Kraus+2017,Frost+2021} ) and outer radii (100s-1000s~au,  \citealt{Johnston+2015,Beltran+2015}) have been derived.

Testing different formation mechanisms of MYSO binaries requires observing them during or just after the formation process. Massive stars spend a significant part of their adolescence hidden within their dusty cocoon and gas clouds. While embedded in their envelope (resulting in a large extinction), some critical but short-lived evolutionary processes are difficult to observe \citep{Zinnecker+2007,Tan+2014,Motte+2018}. Combined with the rarity of such objects, obtaining high quality observational constraints remain challenging. 

The giant \Hii region M17 (also known as the Omega Nebula), is one of the youngest \citep[$\lesssim$1~Myr\; ][]{Hanson+1997}, closest ($\sim$1.7~kpc, \citealt{Kuhn+2018}), and most luminous (L = $3.6\times10^6$\Lsun, \citealt{Povich+2007}) \Hii regions in the Milky Way. 
In a previous spectroscopic study probing the multiplicity of the youngest stars in the super-bubble, \citet{Ramirez+2017} showed that the observed radial velocity (RV) dispersion in M17 is very low ($\sigma_\mathrm{RV}\approx 5$~\kms). This is not compatible with the presence of short-period binary systems that are typically found in OB star populations \citep{Sana+2006, Sana+2012, Kiminki+2012, Sana+2013, Dunstall+2015}. This suggests that either the outcome of massive star formation in M17 has resulted in a significantly lower binary population or that massive binaries are originally formed at larger separations \citep[100~\Rsun\ or more; ][]{Sana+2017}. The latter authors argue that the first explanation is unlikely and propose that the newly born long-period binary systems should migrate on a time-scale of the order of $\sim$2~Myr or less to match the observational properties of main-sequence massive star populations of only a few Myr of age. In the context of disk fragmentation theories, such an inward migration process may be driven by the interaction with the remnant of the accretion disk or with other protostellar bodies of the cluster. Effects of migration on the binary period distribution will halt when the sinks of angular momentum disappear, via the dissipation of the disk or dispersal of the cluster, or as other protostellar bodies are pushed far out or are ejected. Additional evidence  for this migration scenario was found by \citet{Ramirez+2021} who show that the observed RV dispersion of stellar clusters positively correlate with their age.

As the \citet{Sana+2017} scenario presumes that binaries are formed at large separations and then migrate, a strong test is determining the presence of a significant number of relatively massive companions at separations corresponding to the expected size of the accretion disk (10-1000~au, corresponding to 6-600~mas given the 1.7~kpc distance to M17). Such separations lie beyond the detection capabilities of spectroscopic campaigns, but can be probed with high-angular resolution observations \citep{Sana+2017b}. In addition, the youngest massive stars are strongly reddened and their spectral energy distribution peaks in the mid to far-infrared. Fortunately the advent of the GRAVITY instrument \citep{GravityCollab+2017} and the large collective areas of the 8.2m Unit Telescopes (UTs) at the ESO Very Large Telescope and Interferometer (VLTI) offer sufficient angular resolution and sensitivity to probe part of the M17 massive star population. In this paper, we present GRAVITY observations of a small sample of presumably single massive stars in M17 and we show that all the objects are indeed part of a multiple system.

 This paper is organised as follows. Section~\ref{section:Obs} presents the observations and interferometric modelling. Section~\ref{section:Results}  describe the  multiplicity status of each object in the sample. We discuss our results in Sect.~\ref{section:Discussion} and conclude in Sect.~\ref{section:Conclusion}.


\section{Observations and Data Analysis}
\label{section:Obs}

\subsection{Target sample and GRAVITY observations}
\label{section:data}

\begin{table*}[!t]
\caption{Journal of GRAVITY observations.}
    \label{tab:Gravity_journal}
    \centering
    \begin{tabular}{llllllccc}
    \hline
    \hline  
    \multicolumn{2}{c}{Object identifier} & MJD & $\alpha$ (J2000) & $\delta$ (J2000) & $K_\mathrm{S}$ & Spectral  & Calibrator & Seeing\\
   \citet{Bumgardner1992}  & CEN (OI) &  & (h m s) & (\degr\ \arcmin\ \arcsec) &  mag & Type & & (\arcsec)\\
    \hline 
    NGC6618-B189SW  & CEN~1a & 58295.295  & 18 20 29.86 & $-$16 10 44    & 6.9\tablefootmark{a} & O4V\tablefootmark{b}   & HD167334    & 0.57  \\ 
    NGC6618-B189NE  & CEN~1b & 58683.226  & 18 20 29.89 & $-$16 10 46    & 6.9\tablefootmark{a} & O4V\tablefootmark{b}   & HD167334  & 0.56    \\
    NGC6618-B111    & CEN~2  & 58239.288   & 18 20 34.49 & $-$16 10 11.84 & 7.5 & O4.5V\tablefootmark{c} & HD 164357   & 0.81  \\
    NGC6618-B98     & CEN~3  & 58239.407    & 18 20 35.39 & $-$16 10 48.58 & 7.7 & O9V\tablefootmark{b}   & HD 164357  & 1.22  \\
    NGC6618-B0      & OI~345  & 58237.405 & 18 20 28.53 & $-$16 13 31    & 7.4 & O6V\tablefootmark{b}   & HD166565   & 0.82   \\
    NGC6618-B260    & CEN~18 & 58239.336    & 18 20 25.87 & $-$16 08 32.40 & 7.8 & O6V\tablefootmark{b}   & HD 164357   & 0.93  \\
    TYC 6265-1977-1 & $-$   & 58573.334 & 18 20 53.85 & $-$16 03 06.42 & 8.3 & G...\tablefootmark{d}  & TYC5689-515-1 & 0.71 \\
    NGC6618-B293    & CEN~7  & 58712.135 & 18 20 24.01 & $-$16 08 18.63 & 9.3 & F8\tablefootmark{d}    & TYC6846-835-1  & 0.72\\
    \hline
\end{tabular} 
\tablefoot{
The first column lists the the objects number following \citet{Bumgardner1992} nomenclature. The second columns lists alternative IDs, the CEN identifiers \citep{Chini+1980}. The third column provides the observation date. The fourth and fifth columns give the right ascension $\alpha$ and declination $\delta$ (J2000).  The 2MASS $K$s magnitude \citet{Hoffmeister+2008} and spectral types are following. The second to last column indicate the calibrator star used, while the last column provides the seeing of the single-epoch observations. \\
\tablefoottext{a}{Combined magnitude of CEN~1a and 1b;}
\tablefoottext{b}{\citet{Ramirez+2017};}
\tablefoottext{c}{\citet{Hoffmeister+2008};}
\tablefoottext{d}{\citet{Broos+2007}.}
}
\end{table*}

We used the VLTI/GRAVITY $K$-band interferometric instrument and the 8.2m UTs to obtain high-angular resolution data of young massive stars in the M17 region. Our sample was designed based on the spectroscopic study of \citep{Ramirez+2017}, complemented with additional known O stars in the region, such as the Kleinmann Star at M17's core \citep{Kleinmann1973}. We applied a strict magnitude cut at $K<10$ given GRAVITY's advertised limiting magnitude in ESO period 101. This resulted in a sample of 21 objects comprising 11 highly reddened sources. Less than half of the original sample was observed as a result of technical problems and poor weather conditions, including none of the highly reddened targets. This paper presents GRAVITY observations of eight optically visible targets that were successfully observed between May 1st, 2018 and July 19th, 2019 using the optical MACAO system for adaptive optics correction in order to optimise the fibre injection. The observations were made in single-field on-axis mode: the light from the same star is split 50\%-50\% to the fringe tracker (FT) channel and science (SC) channel.

\begin{figure}[!t]
\centering
\includegraphics[scale=0.3]{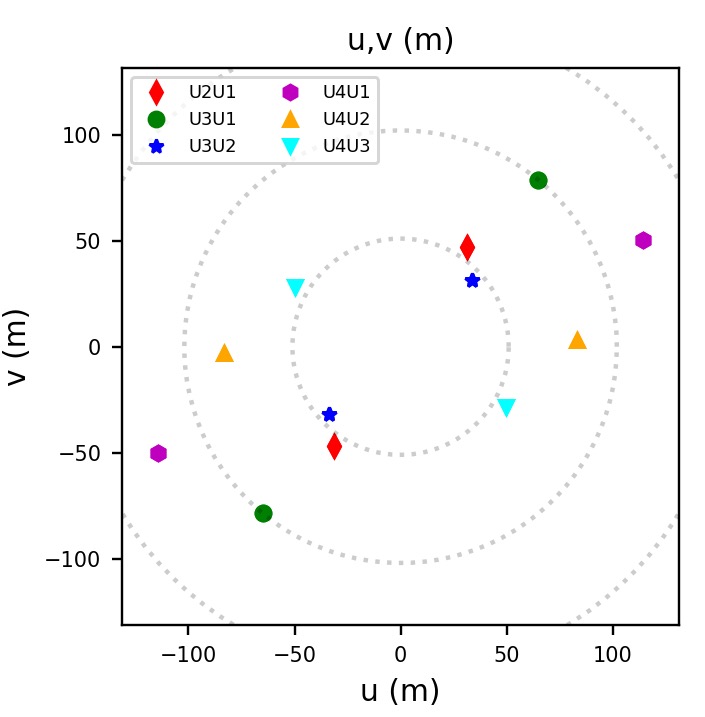}
\caption{Example of projected baselines (so called $uv$-plane) of the VLTI/GRAVITY+UTs observations towards NGC6618-B189SW. The $uv$-coverage slightly differ for the other sources but remains roughly the same due to close sky positions.}
\label{fig:uvPlane}
\end{figure}

The four-UT configuration (U1-U2-U3-U4) delivers baselines ranging from $\sim$50 to $\sim$130~\si{\meter} (Fig. \ref{fig:uvPlane}). The projected baseline lengths are equivalent to a resolution ranging from $\sim1.6$ to 4.3~mas, at the wavelength of 2.16~\si{\micro\meter}. GRAVITY coherently interferes the collected light with double-mode fibres for fringe tracking on one side, and observing science targets on the other side, essentially meaning GRAVITY is both a fringe tracker and science instrument combined in one. The observations were carried out at medium spectral resolving power ($\lambda / \Delta\lambda\sim500$) with individual integration times of 5-30~s, providing a sufficient resolution to perform basic spectral analysis of the observed systems, for instance deriving the stellar parameters as explained in Sect. \ref{section:Discussion}. Table~\ref{tab:Gravity_journal} provides the journal of our observations.

We reduced the data using the standard GRAVITY pipeline \citep{Lapeyrere+2014} and the Python reduction tools from the GRAVITY consortium. To correctly remove the atmospheric effects and the instrumental instabilities during the processing, standard calibrators are observed concurrently. Calibrating the visibilities in order to get the intrinsic visibility of the scientific target is crucial. The choice of the most suitable calibrator was made using the Jean-Marie Mariotti Center (JMMC) SearchCal \citep{Bonneau+2011}. These calibrators are early-K~III stars, for which the known diameters ($<0.4$mas) indicate  that they can be considered as unresolved, single stars.
The visibility calibration was done thanks to the \emph{viscal} pipeline available to the GRAVITY users community via the EsoReflex\footnote{http://www.eso.org/sci/software/esoreflex/} environment.The entire reduced dataset is available on the Optical interferometry DataBase (\texttt{OiDB}) \footnote{Accessible at http://oidb.jmmc.fr} under the collection name \emph{M17 young O-stars}.

\subsection{Distance considerations}
\label{section:distance}

The distance towards M17 has a linear influence on the derived physical separations. Literature estimates vary from 1.3 to 2.9~kpc. Using multicolour UBVRI photometry,  \citet{Chini+1980}  derived a photometric distance of $2.2\pm0.2$~kpc. \citet{Nielbock+2001}  used the same UBV photometry method as \citet{Chini+1980} but reconsidered the extinction law and eliminated the objects with IR excess, yielding $1.6\pm0.3$~kpc. By reconstructing the SED of M17 from near-IR to radio continuum, \citet{Povich+2007} used the integrated flux to derive the bolometric luminosity of $2.4\times10^{6} L_{\odot}$ and thus a luminosity distance of $1.6{\substack{+0.3 \\ -0.1}}$~kpc, in excellent agreement with \citet{Nielbock+2001}. Yet, proper motion of methanol masers in the region provided a slightly larger distance of $d=1.98{\substack{+0.14 \\ -0.12}}$~kpc \citep{Xu+2011}. Maser distances are usually considered very robust and the value of $\sim$2~kpc was therefore adopted in earlier papers by our team \citep{Ramirez+2017, Sana+2017}. 

As an independent check, we use the GAIA DR2 catalogue of distances provided by \citet{Bailer-Jones+2018} and build the distance generalised histogram of all Gaia sources within a 5\arcmin-radius from M17's centre. We repeat the same experiment while restraining the sample to our initial targets list described in Sect.~\ref{section:data}. In both cases, the histograms peak at distances of about 1.5 and 1.6~kpc, respectively. The latest release of Gaia eDR3 gives us an equivalent distance of 1.59~kpc.Both are in good agreement with the more careful work of \citet{Kuhn+2018} whose method supersedes ours and whose paper is fully dedicated to the measurement of distances. Indeed, the authors used the weighted median of individual stellar parallax measurements to estimate the overall system parallax. Field-star contaminants are discarded and a total of 82 stars served for the calculation of M17 distance. The uncertainties are evaluated using bootstrap analysis, to which they added a systematic uncertainty that accounts for a noise floor. In the following, we instead adopt the latter results of \citet{Kuhn+2018}, hence a distance of $1.68{\substack{+0.13 \\ -0.11}}$~kpc.

\subsection{Interferometric modelling}
\label{section:DataAnalysis}

 From the reduced data, we determine the binary parameters by fitting two main types of observables: the visibility amplitude (hereafter |V|) and the closure phase (hereafter CP). The modelling functions and tools used for data analysis are described in this section. 
 
From |V| and CP, we can derive the variable parameters of flux ratio ($f$) and the source positions $(\Delta\alpha ,\Delta\beta)$ by performing binary model fitting. Two main tools have been used for analysing the data: \texttt{LITpro} \citep{Tallon-Bosc+2008} and PMOIRED\footnote{https://github.com/amerand/PMOIRED}, that were developed by the JMMC\footnote{http://www.jmmc.fr/} collaboration and Antoine Mérand respectively. Based on a Levenberg-Marquardt algorithm, \texttt{LITpro} is an useful tool to search for the geometrical functions that best models our observations. In the fitting process, we iteratively select one (single star), two (a binary) or three (a triple) unresolved sources. In order to slightly adjust or improve the fit, we used additional model complexity such as a background flux, or a circumstellar disk. The flux weights of the primary star and the companion were free parameters. The position of the primary star was set to $(0,0)$ at the centre, the position of the secondary and other components were free. \texttt{LITpro} computes $\chi^{2}$ maps in order to accurately set the parameters. Then, a $\chi^{2}$ minimisation is performed in order to find the best-fitting parameters set.

\texttt{LITpro} allowed us to obtain good estimates of the binary position and average flux ratio, but does not allow to explore the values of parameters using a grid nor to handle the spectroscopic aspect of the GRAVITY data. We therefore used PMOIRED, a Python 3 module, which extends the fitting in the spectral dimension and does a telluric correction of the spectrum.
The best-fit model is found throughout a comprehensive fitting process. First, we check the relevance of a single star model, and we iteratively add complexity (binary, triple, resolved component). The uniform diameters of the two (or three) stars were fixed to 0.2~mas so that they are unresolved in our GRAVITY data. Similarly, the $x$ and $y$ spatial positions of the primary star were fixed at coordinates $(0,0)$. For each star the basic morphology of the source has been identified with \texttt{LITpro}. As such, these parameters are used as first guesses for flux and position $(x,y)$. As we wish to characterise any circumstellar emission, we also allow the algorithm to account for a fully resolved emission component, as a fraction of the total flux. We fit geometric models to |V| and CP over the entire wavelength range ($\sim$2.05 to 2.45~\si{\micro\meter}). To precisely determine the binary astrometry, we make use of a 2D search grid of global minima. The estimated parameters for the companion position are the starting point of the fit grid to ascertain the most suitable binary vector. We explore systematically the parameter space over a given pattern ($x,y_{min}$; $x,y_{max}$; $x,y_{step}$) to find the global minimum. The choice of spatial step for the first guesses grid is chosen in such a way that the global $\chi^{2}$ minimum within the explored range is not missed. This method was tested and has proven its efficiency for the CANDID code \citep{Gallenne+2015}. A considerable advantage of PMOIRED is also the possibility of refining the estimation of the parameters, by using bootstrapping. This method usually lead to larger but more realistic uncertainties because bootstrapping mitigate the effects of correlated data.

We explored a second approach which consists in modelling |V|, CP and the normalised flux. The module extracts the continuum and uses it to compute the differential phase and normalised spectrum. 
We select a wavelength range from $2.15$ to $2.18$~\si{\micro\meter}, centred on the Br$\gamma$ line ($2.166$~\si{\micro\meter}). Additional parameters are added such as the flux, central wavelength and width of the line, using a Lorentzian profile. The output value of the flux of the line indicates whether the line is in absorption or emission. In all cases, centring the fit on the Br$\gamma$ line did not significantly change the flux ratio, positions and $\chi^{2}$ previously found.

The goodness of fit of the models in both \texttt{LITpro} and PMOIRED is quantified through the use of reduced chi-square ($\chi^{2}_{r}$). 
Increasing the number of parameters in the model can improve the $\chi^{2}_{r}$ value, so we performed an F-test to decide when adding complexity (i.e adding circumstellar or interstellar components) does not significantly improve the fit. The p-value for the F-test is compared to the 5\% significance level. For example, the fit of |V| and CP for B189NE is noticeably improved by adding a fully resolved component to the model (see Appendix \ref{appendix:Fit_parameters}). The binary parameters used in the rest of the paper, and their respective bootstrapped uncertainties, are those estimated with PMOIRED over the wavelength range $2.05-2.45$~\si{\micro\meter}.

\begin{figure*}[!h]
\centering
\includegraphics[scale=0.45]{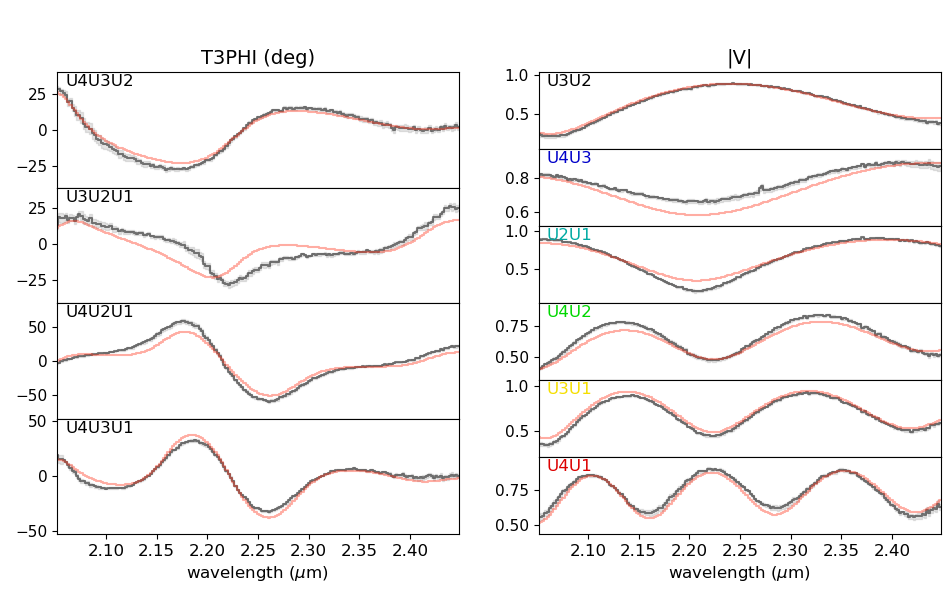}
\caption{The measured visibility amplitudes and closure phases for B189SW, plotted as a function of wavelength for each baseline. The overlaid red lines are the best\-fit model from PMOIRED. The most suitable model for this source is a triple system with most likely a disk around the farthest component. A full description of the model geometry is given in Appendix \ref{tab:B189SW_fit}. }
\label{fig:ex_visamp_CP}
\end{figure*}

\subsection{Spectral fitting and mass estimations}
\label{subsection:spectral_fitting}

The flux spectrum of each object was computed by averaging the four medium-resolution UT spectra stored in the dataset. The absorption telluric features were corrected on both the target and its calibrator using the appropriate module in PMOIRED. The spectrum of the calibrator was used to correct the target from the instrumental response and a K0III standard star allowed to scale the target spectrum. The 2MASS $K$-band magnitude of each star was adopted to flux calibrate the final spectrum. To retrieve the stellar parameters of the binary components, the flux ratios derived for each systems were applied to the spectrum.

In order to constrain the stellar parameters of the targets and their companions, we used the observed calibrated spectra and compared them to family of atmosphere models and evolutionary tracks. 
We combine LTE PHOENIX models ($\leq 3500$~K), ATLAS9 (3500-15000~K) and TLUSTY ($\geq 25000$~K) models to cover the entire range from 2300 to 50000~K \citep{Castelli2003,Lanz+2003,Lanz+2007}.

\begin{figure}[!t]
\centering
\includegraphics[scale=0.65]{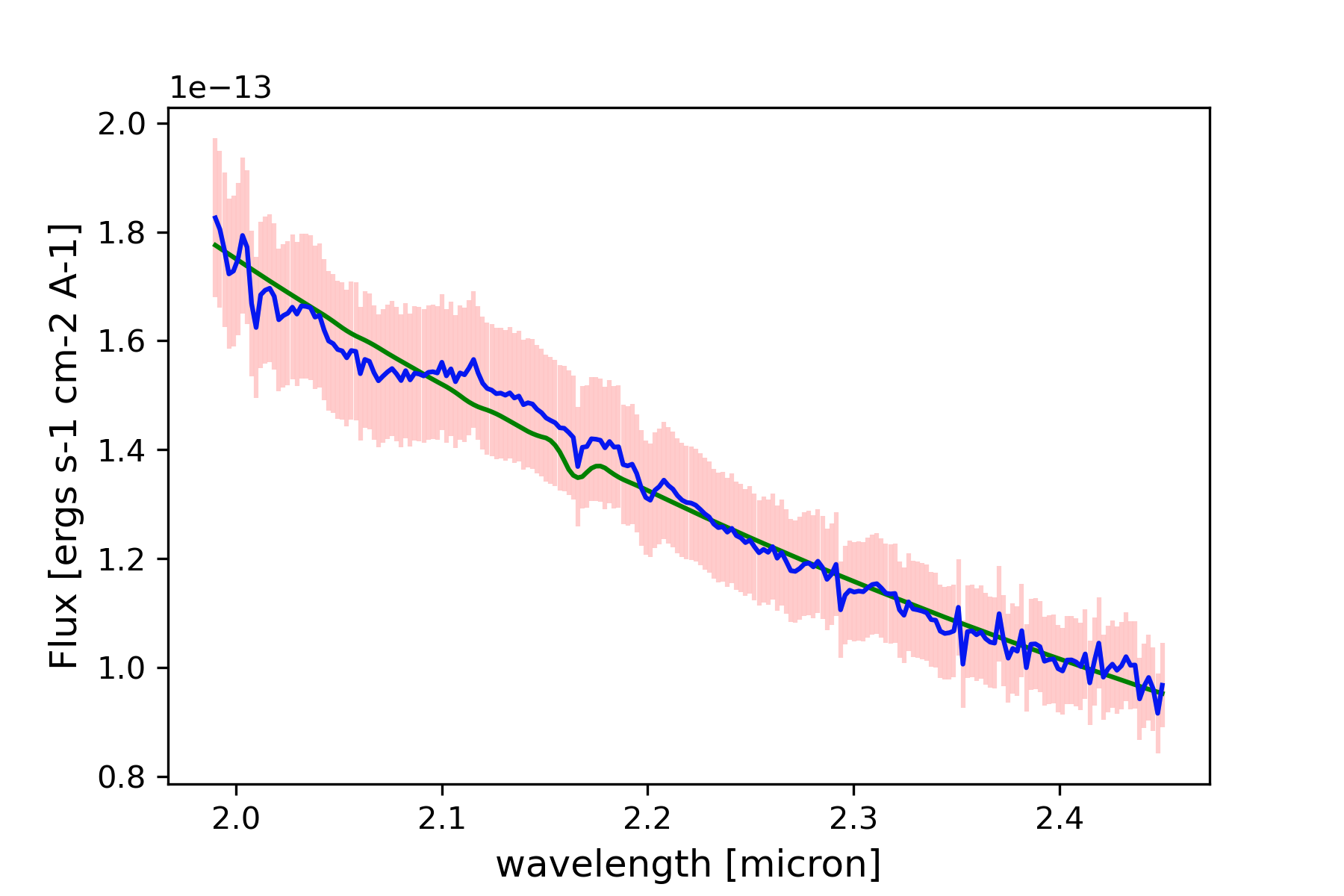}
\caption{Calibrated spectrum of the B189 SW primary star (blue), with errors in red. The best-fit using \citet{Brott+2011} evolutionary tracks is plotted in green.}
\label{fig:spectrum}
\end{figure}

We further adopt the (pre)-main sequence ((P)MS) evolutionary tracks of \citet{Siess+2000} up to 7~\Msun. Above 7~\Msun, we consider the stars to have reached the main sequence and we use the galactic grid of evolutionary tracks and isochrones from \cite{Brott+2011}.
For each mass and age value in the tracks, we linearly interpolate over the atmosphere model grids to calculate the corresponding atmospheric model. We rebin it to the wavelength range and we convolve it to the spectral resolution of our GRAVITY data. We then redden it according to the measured values of extinction for each object \citep{Hanson+1997,Hoffmeister+2008,Ramirez+2017} and the \cite{Fitzpatrick+2019} reddening law. 
Finally, we compare models and data through a $\chi^{2}-$metric, taking into account the uncertainty on the observed spectrum. An example of a spectrum fit is shown in Fig. \ref{fig:spectrum}.

Within the 68\% confidence interval, several combinations of stellar parameters are consistent with the observations. For our best fits, we choose those with effective temperature that best matches the known spectral type of the primary. Ages are also taken into account given the young age of M17 ($\lesssim$1~Myr). Once the mass of the primary is determined, the magnitude contrast helps to gauge the difference in the sub-spectral type between a star and its companion, hence its approximate mass. The same process is then iteratively used to look for the best fit model of the companion.
Ultimately, in view of the derived interferometric parameters, all systems appear to be massive. In this context, fits combining the ATLAS or TLUSTY atmospheric models with the \cite{Brott+2011} evolutionary tracks are the most relevant. \text{red}{} All the masses, for both central objects and companions are displayed in Appendix \ref{appendix:Master_Table}. 

\section{Multiplicity result}
\label{section:Results}

\begin{figure}[!t]
\centering
\includegraphics[scale=0.65]{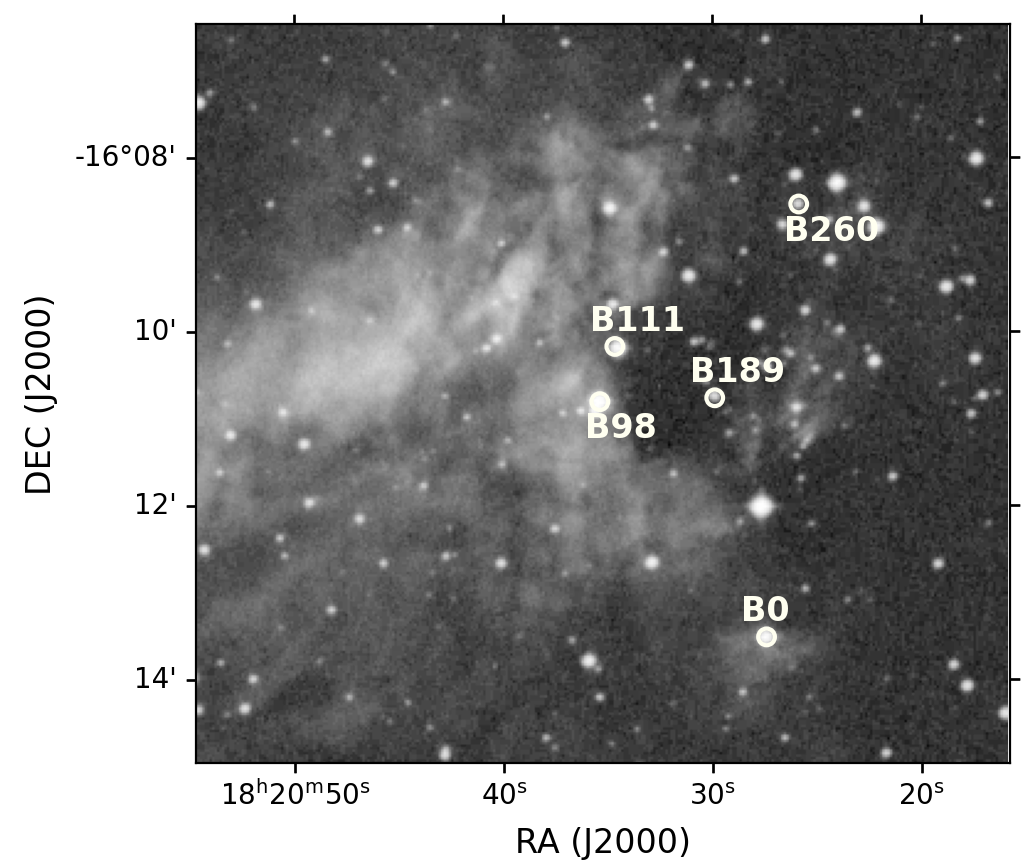}
\caption{DSS image of NGC6618 on which we labelled our M17 GRAVITY sources in white.}
\label{fig:M17sources}
\end{figure}

We summarise below the interferometric results for each target in our sample. In this section, we first concentrate on the six stars that belong to the NGC6618 complex (Fig. \ref{fig:M17sources}). At the end of this section we present two more systems present in our data but which, after checking their distance \citep{GaiaDR2}, proved to be foreground stars. With long-baseline interferometry we reveal companions at relatively small angular separations, ranging from $\sim$2~au to 100~au given a typical distance of 1.7~kpc to M17. For reference, these angular separations correspond to intermediate orbital periods of about 2.3 $\lesssim$ log$P$(days) $\lesssim$ 4.3. This is a rough estimation of the period that strongly depends on the eccentricity, that we cannot derive from our single epoch GRAVITY data. In the following, we summarise our results for each object. The best-fit models and a table with the fitted parameters are respectively presented in Appendix \ref{appendix:GRAVITY_DATA} and \ref{appendix:Fit_parameters}. The latter appendix also shows how the $\chi^{2}_{r}$ varies with respect to the models. 

\subsection{NGC6618-B189}
\label{subsubsection:B189}

Also known as the Kleinmann's Anonymous Star, NGC6618-B189 \citep[hereafter CEN1 or B189]{Kleinmann1973,Chini+1980} is a massive visual binary in the centre of NGC6618 and the most luminous object in M17. It was thought to be composed of two highly reddened  O4 stars,  with a separation of $\sim$1.8$"$ ($A_\mathrm{V}$=10 and 13 mag, \citealt{Hoffmeister+2008}).  The two components, that  are major ionising sources in the region, were resolved by  CHANDRA and XMM-Newton as bright  X-ray  sources with a hard thermal plasma emission \citep{Broos+2007, Mernier+2013}. As a result, these authors suggested that the O4+O4 system actually consists of 4 massive stars, i.e. two pairs of spectroscopic binaries featuring strong colliding winds or magnetic properties. 
\citet{Broos+2007} reported a total mass of $\sim$140~\Msun\ for the combined O4+O4 system. The Kleinmann Star has also been observed in the radio regime. \citet{Rodriguez+2009} outlined the presence of two radio sources matching the optical binary CEN 1. The radio emission likely originates from synchrotron emission produced by a relativistic population of electrons accelerated by shocks in the vicinity of the wind-wind collision regime. 

Our interferometric observations demonstrate that NGC6618-B189 is a binary+triple system. We discuss separately the results for each component as the GRAVITY observations of B189 were split into two different pointings in order to characterise each component separately:  NGC6618-B189NE \citep[also known as B189NE or CEN1a,][]{Hoffmeister+2008} and NGC6618-B189SW \citep[also knwon as B189SW or CEN1b,][]{Hoffmeister+2008} were observed one year apart, in June 2018 and July 2019, respectively.

\begin{figure*}[!h]
\centering
\includegraphics[scale=0.58]{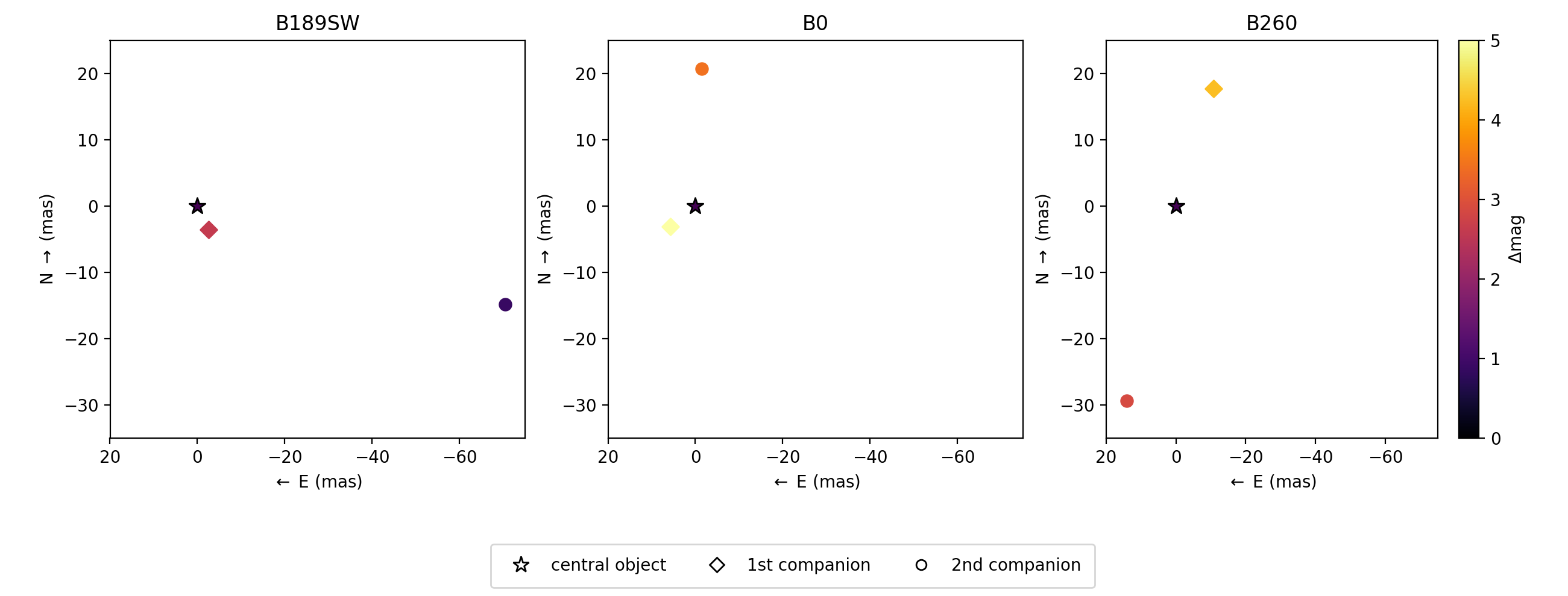}
\caption{Configuration of the three triple systems B189SW (MJD: 58295.295), B0 (MJD: 58237.405) and B260 (MJD: 58239.336). The companion stars are colour scaled according to their respective flux ratio (in delta magnitude scale). The flux of the central object is set at 1 by default.}
\label{fig:config_triples}
\end{figure*}

\textbf{B189SW} is also reported as an O4 object.  
This system is very intriguing, due to the unusual shapes of the visibilities and closure phases, as shown in Fig. \ref{fig:ex_visamp_CP}. 
The system is characterised by a high-frequency sinusoidal shape in the visibilities, indicating multiple stars at large separations. Similarly, the closure phases oscillate with non-constant amplitudes, between $-$60 and $+$60 degrees at the most, showing the high asymmetry of the system. B189SW consists of at least three hierarchical components (see Fig.~\ref{fig:config_triples}). The shapes of |V| and CPs are best reproduced using a triple system. In addition, the visibilities do not reach 1 when extrapolating to $B=0$, indicating the probable presence of a resolved component. In order to better recreate the visibility modulation and obtain the correct amplitude on some baselines, the farthest companion is set as a deformed stretched Gaussian, whose direction is characterised by two additional parameters, the projection angle and the inclination.
The first detected companion is described as a resolved object with a flux of $44 \pm 4$\%\ and a spectral index of $-2.4\pm 0.5$, a Gaussian FWHM of $3.97\pm0.41$~mas oriented along a projection angle of $-42.8\pm3.6^{\circ}$, an inclination of $77.7\pm5.9^{\circ}$ and has an angular separation of $\rho = 71.95\pm2.49$~mas ($d = 120.88\pm 4.18$ $(\pm ^{+9.35}_{-7.9}$)~au, \textbf{\citealt{Hummel+2016}}).

The second detected companion is closer ($\rho =4.40\pm0.82$~mas; $d = 7.40\pm 1.37 \pm ^{+0.57}_{-0.48}$~au) and is much fainter than the central object $F_{3}=0.09 \pm 0.02 F_{1}$. The complex geometry of the system is presented in Appendix \ref{appendix:Fit_parameters}.

With almost two magnitude differences in the $K$-band between the brightest object of the system and the other two companions, the latter are likely late-O or early-B stars. In such circumstances, and given the quite large separations involved, it is more difficult to justify the high X-ray luminosity and hard spectrum of B189SW from wind-wind collision. Indeed the late-O or early-B stars have much weaker winds than the O4 star, so that either the wind collision occurs close to the surface and wrap around the weaker wind stars, or even collapse on the stars' surfaces \citep{Usov+1992,Sana+2005}. In either case, the geometrical cross-section will be small, hence should not lead to strong X-ray over luminosity. An alternative explanation may be that yet another massive companion exists near the brightest component that cannot be resolved by the $\sim$2~mas angular resolution of our GRAVITY data, or that the X-ray emission from B189SW has a magnetic origin \citep{UsovMelrose+1992}.

\textbf{B189NE} is known as an O4 star and one of the brightest and most massive members of M17 ($m_\mathrm{K}=5.8$). The visibility and closure phase pattern are in good agreement with a binary model and a fully resolved component that is likely a disk. As we do not probe baselines short enough, the exact nature and geometry of the resolved component cannot be determined. However, the parameters of the binary do not depend on the choice of the resolved component, but do change when we introduce the resolved component. Omitting a resolved component biases the astrometry and flux ratio of the binary, independently of its nature. We detect an interferometric companion at a separation of $\rho=1.26\pm 0.02$ mas, corresponding to a projected distance $d$ of $2.12 \pm 0.03 (\pm ^{+0.16}_{-0.14}$)~au, where the first uncertainty is that of the interferometric measurements and the second corresponds to that on the distance. Our best-fit model attributes 15\% of the total emission to the fully resolved component. We measure a $K$-band flux ratio of $f=F_\mathrm{2}/F_\mathrm{1}=0.75\pm 0.01$ between the two stars which, according to the calibrations of \citet{MartinsPlez2006}, translates into a difference of about one spectral sub-type between the two components of B189NE. Such a pair of early O stars would have strong stellar winds, and indeed are expected to generate a large amount of X-ray emission through wind-wind collision \citep{Usov+1992,Stevens+1992,Sana+2004,Sana+2006}. 

\subsection{NGC6618-B111}

NGC6618-B111 (also known as B111, CEN 2 or OI 337) has been reported to be an O5~V source and has a $K$-band magnitude of $m_\mathrm{K}= 7.47$ \citep{Hoffmeister+2008}. The spectral type of B111 has been recently revised thanks to an {\it X-SHOOTER} campaign. Referring to the strong \ion{He}{ii} lines, \citet{Ramirez+2017} classified B111 as a O4.5~V star. Despite the young age of B111, \citet{Hanson+1997} did not report any disk signature nor any NIR excess. 

Based on the non-LTE stellar model FASTWIND \citep{Puls+2005,Rivero+2012}, \citet{Ramirez+2017} derived stellar properties such as the effective temperature ($T_\mathrm{eff}=42850{\substack{+3050 \\ -1600}}K$) and the spectroscopic mass ($M_\mathrm{spec}=42{\substack{+12 \\ -13}}$\Msun). Such small flux ratio has probably but a limited impact on the spectroscopic results of \citet{Ramirez+2017}.

With GRAVITY, we detect a companion with a flux ratio $f=0.03\pm0.01$ at a separation of $17.26\pm0.97$~mas ($d=29.00\pm1.64\pm^{+2.24}_{-1.90}$~au). The data are best fitted with adding a a resolved flux component, that contributes for $5.0\pm0.1~\%$ to the total flux.

\subsection{NGC6618-B98} NGC6618-B98 (otherwise known as CEN3 and referred to hereafter as B98) is a O9V-type source with a $K$-band magnitude $m_\mathrm{K}= 7.58$. Using {\it JHK} imaging at $1.1"$ angular resolution, \citet{Hoffmeister+2008} barely resolved B98 into two equally bright components in the $K$-band. They also reported a third very red companion, at 2.5" in the southeast direction. This third companion is 0.7 mag fainter than the central binary. \citet{Povich+2009} note the low X-ray emission of B98 and pointed out the possibility of a companion.  
We used GRAVITY to further investigate the close environment of B98. 
Using our GRAVITY data we detect a new companion at $1.01\pm 0.03$ mas ($d=1.70\pm0.06\pm^{+0.13}_{-0.11}$~au), with an average flux ratio $f=0.31\pm 0.03$. 
 
\subsection{NGC6618-B0}
NGC6618-B0 (otherwise known as B0 or OI~345) is an O6~V source \citep{Hoffmeister+2008} with a $K-$band magnitude of 7.48. No companion has been reported in previous literature. With GRAVITY snapshot, we can either fit a binary or a triple model, both with an equivalent $\chi^{2}_{r}$ of respectively 1.75 and 1.63 (see Appendix \ref{appendix:Fit_parameters}). The first resolved companion lies at a separation of $20.85\pm1.25$~mas ($d=35.022\pm2.10\pm^{+2.71}_{-2.29}$~au) with a flux ratio $f=0.04\pm0.01$. We eventually detect a second companion with a flux ratio $f=0.010\pm0.005$ at a separation of $6.50\pm0.75$~mas ($d=10.84\pm1.26\pm^{+0.84}_{-0.71}$~au).  Given its flux ratio, the second companion is likely an early B-type star while the first one is a less massive object, probably  still on a pre-main sequence track given the young age of M17. The slight difference in the $\chi^{2}_{r}$ and in the number of degrees of freedom does not allow an accurate determination of the nature of the system. By performing an F-test with a significance level fixed at 5\%, we derive a p$-$value of 0.0199, showing evidence towards a triple system. However, this is not further confirmed if we lower the significance level to 1\%. In the following discussion, we treat B0 as a triple system (as represented in Fig.~\ref{fig:config_triples}), but we clarify in Sect. \ref{section:Discussion} how this choice affects the companion fraction. However, getting additional interferometric data and performing image reconstruction may help us to image the close environment of B0, and as such, better constrain the multiplicity of this object.

\subsection{NGC6618-B260}
 
NGC6618-B260 (or CEN 18, hereafter B260) is an O6~V source, with a $K$-band magnitude of $m_\mathrm{K}= 5.8$. Displaying double absorption lines, in particular showing equally bright double Pa-11 lines, B260 is a candidate double-lined spectroscopic binary \citep{Hoffmeister+2008}. The GRAVITY visibility amplitudes and closure phases reveal two faint companions (see Fig.~\ref{fig:config_triples}) and the presence of a resolved component that accounts for $9.0\pm0.1~\%$ of the total flux, and for which we cannot constrain the exact nature. The first companion is detected at a separation of $32.55\pm2.45$~mas ($d=56.69\pm4.12\pm^{+4.23}_{-3.58}$~au) with a flux ratio $f=0.07\pm0.01$.
The second companion, closer, orbits at $20.70\pm4.76$~mas ($d=34.76\pm7.99\pm^{+2.69}_{-2.28}$~au) from the inner star and has a flux ratio of $f=0.02\pm0.01$. \textbf{}
Given the measured brightness ratio, these stars are unlikely the same companions as those detected in spectroscopy, making B260 a candidate hierarchical quadruple systems. The inner SB2 pair is likely separated by less than 1~mas as otherwise we would have detected its signature, which suggests probable orbital periods of several months, or less.

\subsection{Nearby stars in the line of sight towards M17}

NGC6618-B293 and TYC6265-1977-1 are F8 and G-type stars respectively. They do not happen to belong to the NGC6618 cluster, nor to M17. \citet{Broos+2007} suggested they could be two foreground stars but we still included them in our magnitude-limited sample as they met our selection criteria (see Sect.~\ref{section:data}) and as GAIA DR1 were not sufficiently robust at the time the latter was built.

\subsubsection{NGC6618-B293}
NGC6618-B293 (or CEN 7) is a F8~V type star with $m_{K}=9.35$. With a Gaia DR2 distance of  at $238\pm4$~pc \citep{Bailer-Jones+2018} (Gaia eDR3: $240\pm1$~pc) , it is much closer than the 1.7~kpc expected for the distance of M17 and a clear foreground star. 
Given a flat visibility at $\sim$1, we conclude that the object is a single star within the sensitivities of our observations.

\subsubsection{TYC6265-1977-1}
 TYC6265-1977-1 (hereafter TYC6265) was claimed to be a G-type star \citep{Broos+2007}, with a $K$-band magnitude of 8.28. TYC6265 is located at $375\pm6$~pc away from the solar system \citep{Bailer-Jones+2018}. Gaia eDR3 measurements are of the same order of magnitude ($379\pm3$~pc) and further confirm that TYC6265 does not belong to M17. The interferometric data reveal the presence of a binary system. A companion with almost the same brightness as the primary is detected at $1.23\pm0.04$~mas ($d=0.46\pm0.01$~au) from the central star, with a flux ratio $f=1.04\pm0.02$. TYC6265 is therefore one of the 50\% of Sun-like stars in the solar neighbourhood to be a multiple system \citep{Raghavan+2010}. In the present case, the situation is rarer as the expected period is only a few months.

\section{Discussion}
\label{section:Discussion}

 Excluding B293 and TYC6265, our M17 sample is comprised of six objects. With up to three companions found per system, detected by spectroscopy or interferometry, all of the objects are multiple systems. In this section we present relevant binary parameters such as the multiplicity and companion fraction, flux and mass ratios, allowing to better understand the multiplicity population in the NGC6618 complex. We also discuss the implication for binary star formation models that could apply to this young star-forming region.

\subsection{Multiplicity and companion fraction}
The immediate quantifiable multiplicity parameters that can be derived from our sample are the frequency of multiple systems (hereafter, $MF$) and the fraction of companions (hereafter, $CF$) \citep{Duchene+2013,Sana+2014}. 

The multiplicity frequency is the ratio of multiple systems to the total number of systems. We derive the uncertainties on the multiplicity frequency using binomial statistics \citep{Sana+2014}. Given the small sample size and the fact that $MF\simeq$100\%, the 68\%-confidence interval is better estimated using Monte Carlo simulations than by using the binomial formula. We generate about 100 000 systems where each systems has a binary fraction of 0.9 \citep{Sana+2012}. We sum the simulated results with 100\% of success and divide them by the number of trials. We repeat the process for a growing binary fraction from 0.9 to 1. For the young O-type stars in M17, we obtain a MF of 100\%, with $MF>94\%$ at the 68\%-confidence interval.

We compare this result with various studies, including those of \citealt{Duchene+2013},\citealt{Sana+2014} and \citealt{GravityCollab+2018}. In their review, \citet{Duchene+2013} present the stellar multiplicity from very low-mass to high-mass stars among main and pre-main sequence populations. Based on different observational techniques, the authors provide the statistics for more than 200 objects in the field from 0.01 to 28~\Msun, with the exception of the high-mass stars ($>$16~\Msun) that are most likely members of open clusters or OB associations. Indeed, only 20\% of the Galactic massive stars are known to be isolated, most of them being runaways \citep{deWit+2004}. The SMaSH+ survey \citep{Sana+2014} obtained high-angular resolution observation of nearly 100 O-type stars spanning different luminosity classes (I-V). It covers separations up to 16 000~au assuming a typical distance of 2~kpc. The vast majority of these stars are found in clusters or associations. For the purposes of this study and to provide a more accurate comparison among O main sequence stars, we restricted the SMaSH+ database to dwarfs, those of Luminosity Class V (matching that of our sources). This leads to a sub-sample of 24 O-dwarfs. \citet{GravityCollab+2018} explored the multiplicity fraction of 16 stars (mainly O and B spectral type) in the young Orion Trapezium cluster, with separations up to 600~au and mass ratios of about 0.3 on average.

\begin{figure}[!t]
\centering
\includegraphics[scale=0.6]{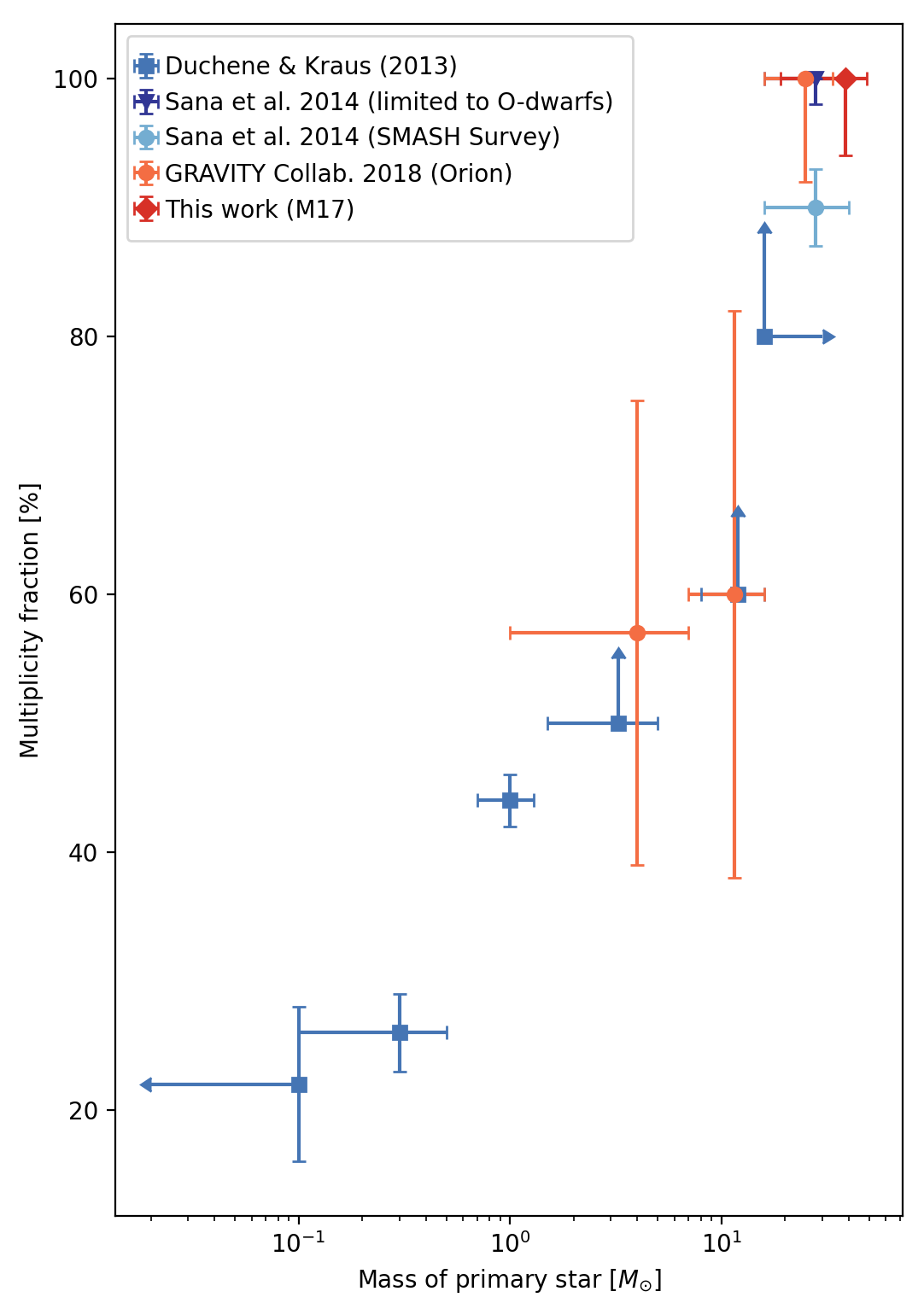}
\caption{Multiplicity fraction ($MF$) of stars in various systems as function of the primary mass. For reference, values of \citet{Duchene+2013} (dark blue), \citet{GravityCollab+2018} (orange) and \citet{Sana+2014} (light blue) are also presented for different mass regimes. }
\label{fig:MF}
\end{figure}

With stars spanning different mass ranges, \citet{GravityCollab+2017} opted for a different strategy by grouping stars according to their mass. In Fig. \ref{fig:MF}, for the mass range $>$ 16\Msun, \citet{GravityCollab+2017} counts 4 sources in the Orion Nebula and obtains a $MF$ of $100{\substack{+0 \\ -9}}\%$ in agreement with what we derive. 
Our results also follow those of \citet{Sana+2014}:
within the sub-sample of O dwarfs described above, they resolved 19 stars with at least one companion closer than 45~mas, hence 79\% of the 24 O dwarfs. However, including previously detected spectroscopic companions, this rate rises to $100{\substack{+0 \\ -5}}\%$ of multiple systems. Fig. \ref{fig:MF} shows a clear trend towards multiple systems for high-mass stars. The 6 systems observed in the M17 star-forming region further attest the multiplicity fractions measured in other galactic regions. Based on the $MF$ parameter, and despite its youth, M17 appears to present a regular massive star population with a 100\% incidence of multiple systems for stars with M$>$20~\Msun.

\begin{figure}[!t]
\centering
\includegraphics[scale=0.6]{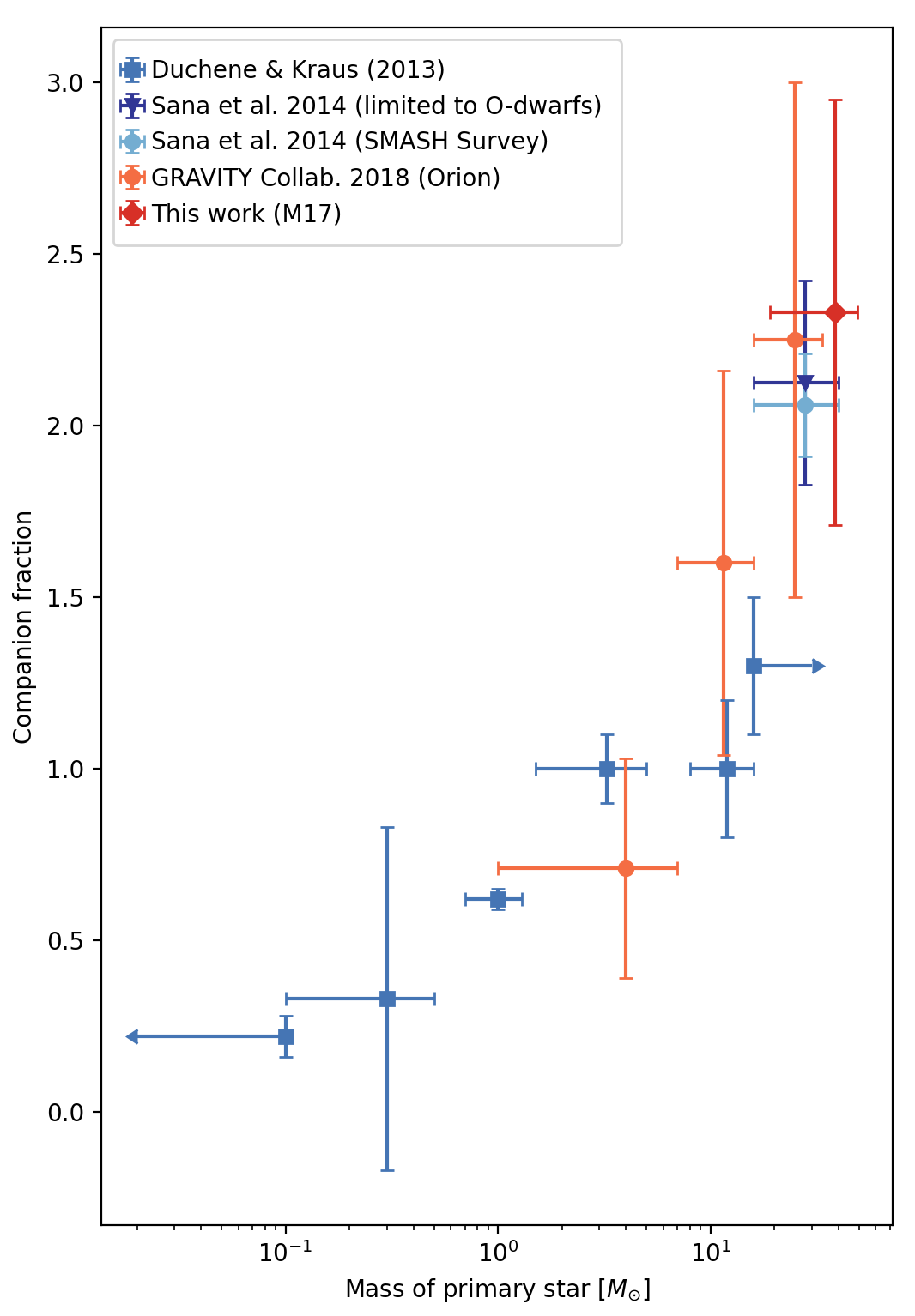}
\caption{Companion fraction (CF) of stars in different systems as a function of the primary mass. For reference, values of \citet{Duchene+2013}, \citet{GravityCollab+2018} and \citet{Sana+2014} are also presented for different mass ranges. The results from \citet{Sana+2014} (SMaSH+ Survey) presented here are those including companions up to 16000~au. Unresolved and resolved companions are taken into account in the final companion rate derived. }
\label{fig:CF}
\end{figure}

The companion fraction is the average number of companions per system, and can thus reach values greater than 1. It is set as the number of companions divided by the total number of systems, including multiple and single stars. We derive the uncertainties on the companion rate using Poisson statistics \citep{Sana+2014}. With 9 companions detected with GRAVITY at separation ranges of 1-100~au, we get a companion fraction of $1.5\pm0.5$. The number of companions per central object can vary from 1 to 3 if we add the spectroscopic companions reported in the literature. With 14 companions for a total of 6 systems, the derived companion fraction reaches $2.3\pm0.6$ (Fig. \ref{fig:CF}), assuming that B0 is a triple. This number is somewhat lowered to $2.2\pm0.6$ insofar as B0 is a binary. The overall conclusion is unaffected given that both CF are within the errorbars.

\begin{figure}[!t]
\centering
\includegraphics[scale=0.6]{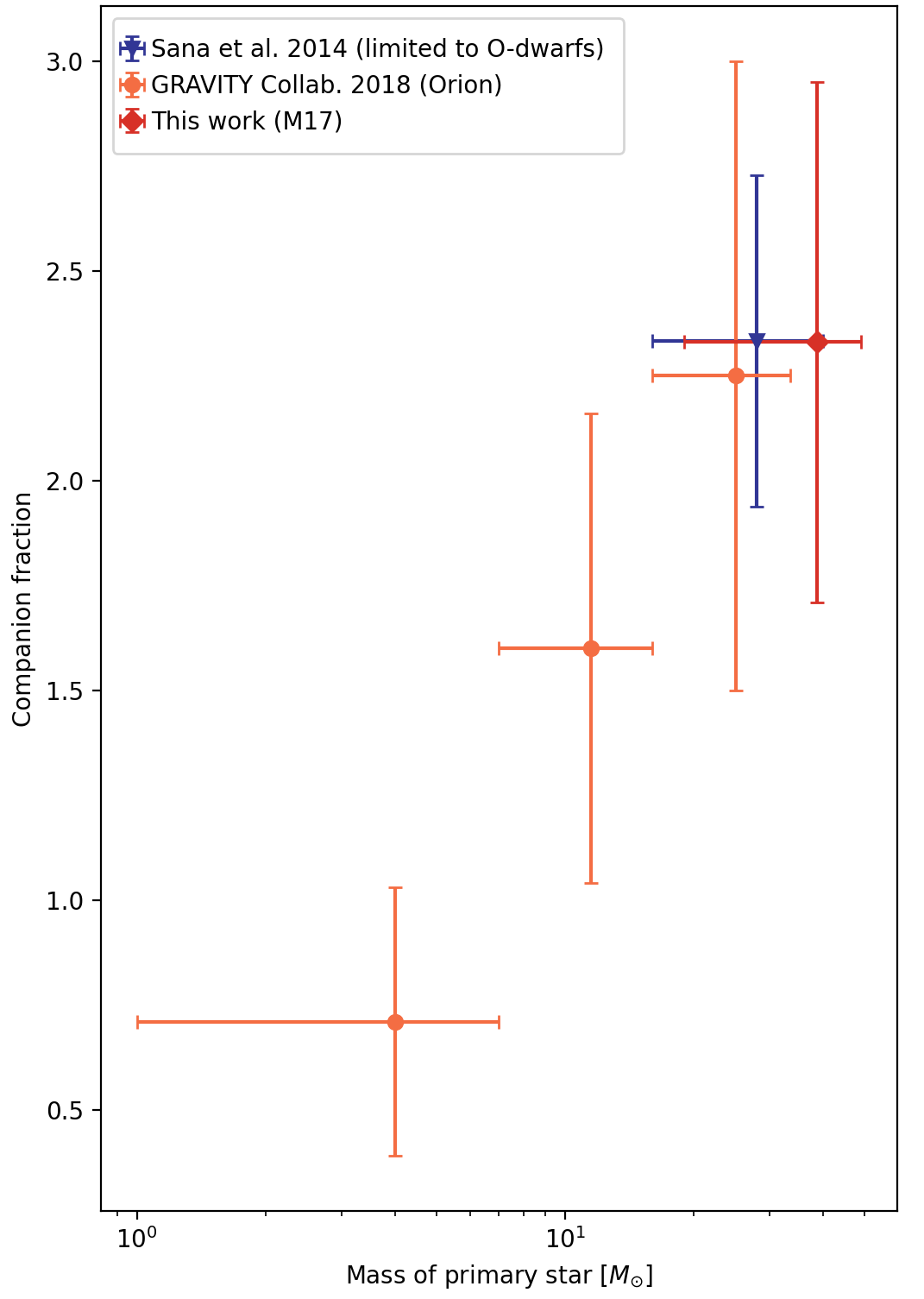}
\caption{Companion fraction (CF) within 100~au in different systems as a function of the primary mass. For reference, values of \citet{GravityCollab+2018} and \citet{Sana+2014} (sub-sample limited to PIONIER observations) are also displayed for different mass ranges.}
\label{fig:CF_100au}
\end{figure}

For the group of stars with M$>$16~\Msun\ (i.e 4 systems), \citet{GravityCollab+2018} observed a companion fraction of $2.3\pm0.8$. This result, includes both spectroscopic and interferometric companions. The results for the two other mass ranges ($1\lesssim$ M $\lesssim7$~\Msun\ and $7\lesssim$ M $\lesssim11$~\Msun) are shown in Fig. \ref{fig:CF}. Still restricting to stars with mass $>$ 16\Msun, \citet{Duchene+2013} noted a companion fraction of $1.3\pm0.2$ lower than the other results by almost a factor of 2. The companion fraction derived by \citep{Sana+2014} depends on the assumed separation range for companions. The full SMaSH+ survey can detect companions up to 16000~au yielding a $CF$ of $2.06\pm0.15$, regardless of the luminosity class. In the case of the restricted selection of SMaSH+ O-dwarfs, the total number of companions is 51 within 600~au, yielding an average $CF$ of $2.13\pm0.30$ \citep{Sana+2014}.
As we cannot probe for such wide companions here, we restricted even more the SMaSH+ database. We picked O-dwarfs observed with PIONIER in the $H$-band, ensuring the detection of companions within 100~au. The new sub-sample is comprised of 15 systems, for which a total of 35 companions have been reported. Fig. \ref{fig:CF_100au} shows the results in a more homogeneous way: all companions fractions are derived within 100~au. We notice that the three datapoints corresponding to O stars show a similar $CF$ despite large errorbars. Based on Poisson statistics, the errorbars could be reduced with an increasing sample size and a more accurate estimation of the spectroscopic companions (some information is lacking within M17 sources). As for the multiplicity fraction, M17 is no exception in terms of companion fraction. The young age of M17 tend to show that massive star formation results in a multiple system, with at least two companions.

\subsection{Flux ratios and mass ratios}
\begin{figure*}[!t]
\centering
\includegraphics[scale=0.55]{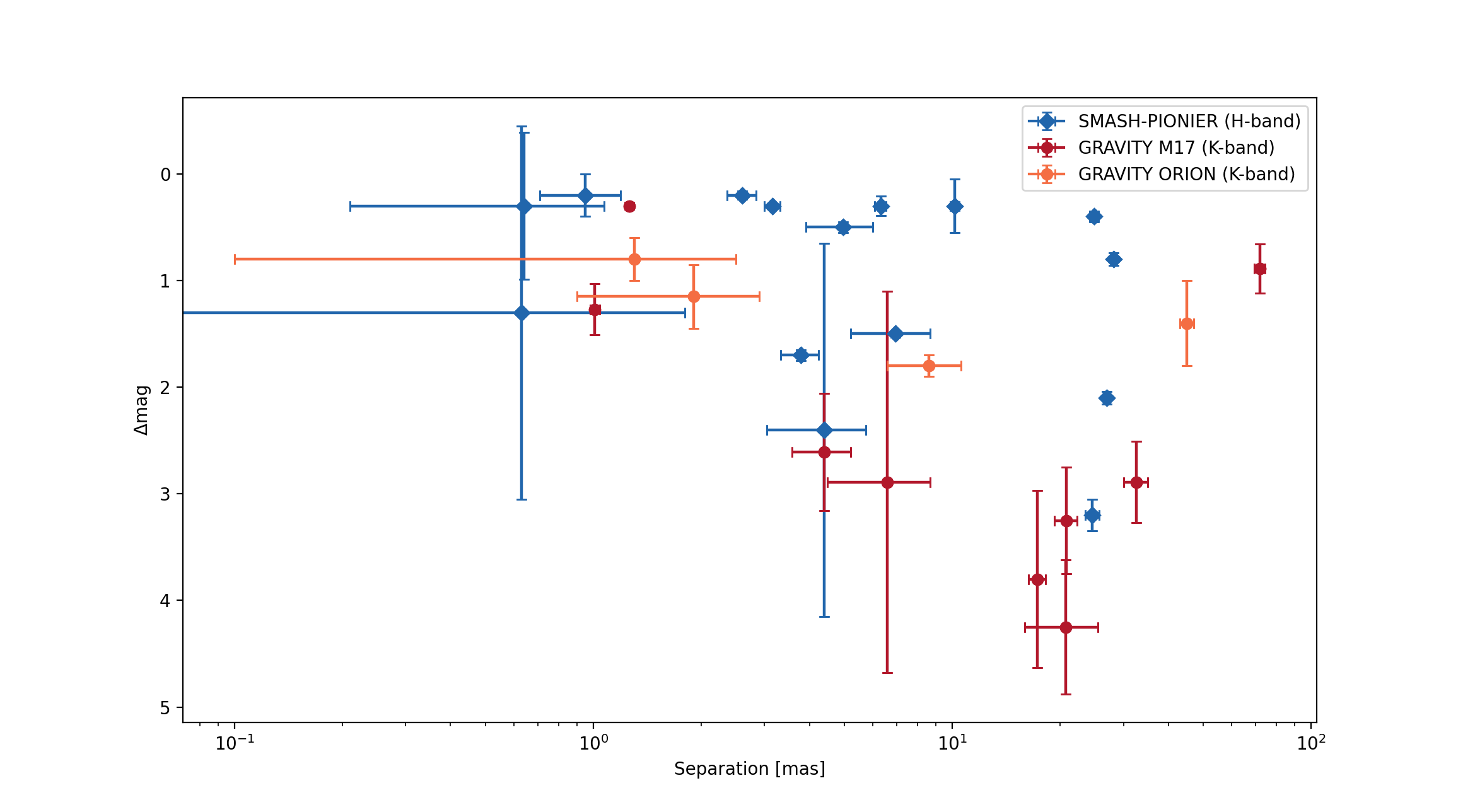}
\caption{Magnitude difference ($\Delta$mag) per angular separation in mas. For reference, the 19 detected companions around O-dwarfs as part of the SMaSH+ survey (blue, \citealt{Sana+2014}) and the 4 resolved companions around high-mass stars in Orion (orange, \citealt{GravityCollab+2018}) have been plotted.}
\label{fig:deltamag_vs_sep}
\end{figure*}

We compare flux ratios, mass ratios and separations of companions by plotting different quantities. In Fig.~\ref{fig:deltamag_vs_sep} we display the resulting brightness contrast as a function of the companion separation in mas, assuming a distance of 1.7~kpc to M17, as described in Sect. \ref{section:distance}. Other interferometric campaigns for the search of binaries have been overlaid, those of \cite{Sana+2014} and \cite{GravityCollab+2018}. Only interferometric companions around O-type stars detected by GRAVITY in the $K$-band (Orion and M17) or PIONIER in the $H$-band (SMaSH+, O dwarfs) are displayed. Only one detection per system has been reported for Orion and SMaSH+ objects. Overall, this plot exhibits the wide range of companion properties that constitute the multiple systems around O-type stars, within a few 100s~mas. No significant relation between the magnitude difference and separation is noticed. There is a clear lack of detection below $\Delta$mag$\sim$2 and companions located within 3~mas, difficult to reach within the PIONIER and GRAVITY sensitivity. SMaSH+ looks at sources with typical distance of 2~kpc, whereas M17 sources are 1.7~kpc away and Orion sources at 414~pc from Earth. The similar brightness binaries observed with PIONIER are lacking in GRAVITY data, for which most of the $\Delta$mag are larger than one. With the exception of B189NE, whose companion which orbits at 2.1~au, is 75\% as bright as the primary. However, looking at M17 companions, we observe larger magnitude difference for companions around 10~mas. The companion of B189SW that lie around 120~au from its host star, is significantly bright with about 44\% of the brightness of the central star. We notice the presence of a companion with similar characteristics (around 50~mas, $\Delta$mag$\sim$1.4) found in the Orion Nebula.

\begin{figure}[!t]
\centering
\includegraphics[scale=0.60]{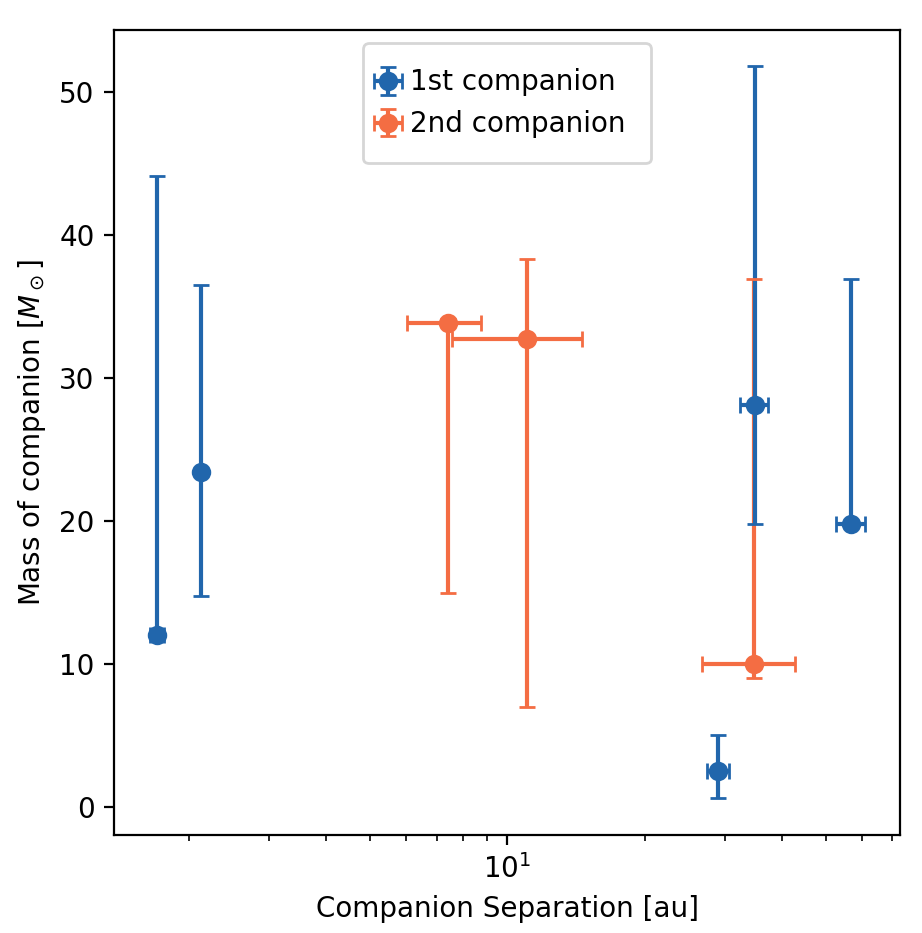}
\caption{Mass of the companions per companion separation (in au). In blue, the first companion refers to the closest companion. In orange, the second companion refers to the second-closest, found with GRAVITY. }
\label{fig:masscomp}
\end{figure}

Figure \ref{fig:masscomp} shows the mass of the companions with respect to their separation. Mass ratios were calculated from the best fit values retrieved after modelling the calibrated spectra (see Sect. \ref{subsection:spectral_fitting}). All the masses, for both central objects and companions are displayed in Appendix \ref{appendix:Master_Table}. We point out that the mass of the farthest component of B189SW has not been derived as its nature may likely involve a disk. As such, the flux ratio might include the contribution of the resolved component and cannot simply be modelled by a low-mass stellar atmosphere model. Consequently, only eight companions will appear in the following plots (out of 9 found). The colours indicate the order of the companion, from innermost (blue) to outermost (orange). There is no clear trend for the companion mass according to the separation. We notice that relatively massive companions are found around the primary stars, reaching up to 49~\Msun, regardless of their proximity to the host star. 
For five companions, the derived masses reach the limits fixed by the limited number of evolutionary tracks in the model. It is notably the case for the closest companion ($\sim$1.7~au) and the farthest companion that lie at large distance (around 120~au).

\begin{figure}[!t]
\centering
\includegraphics[scale=0.60]{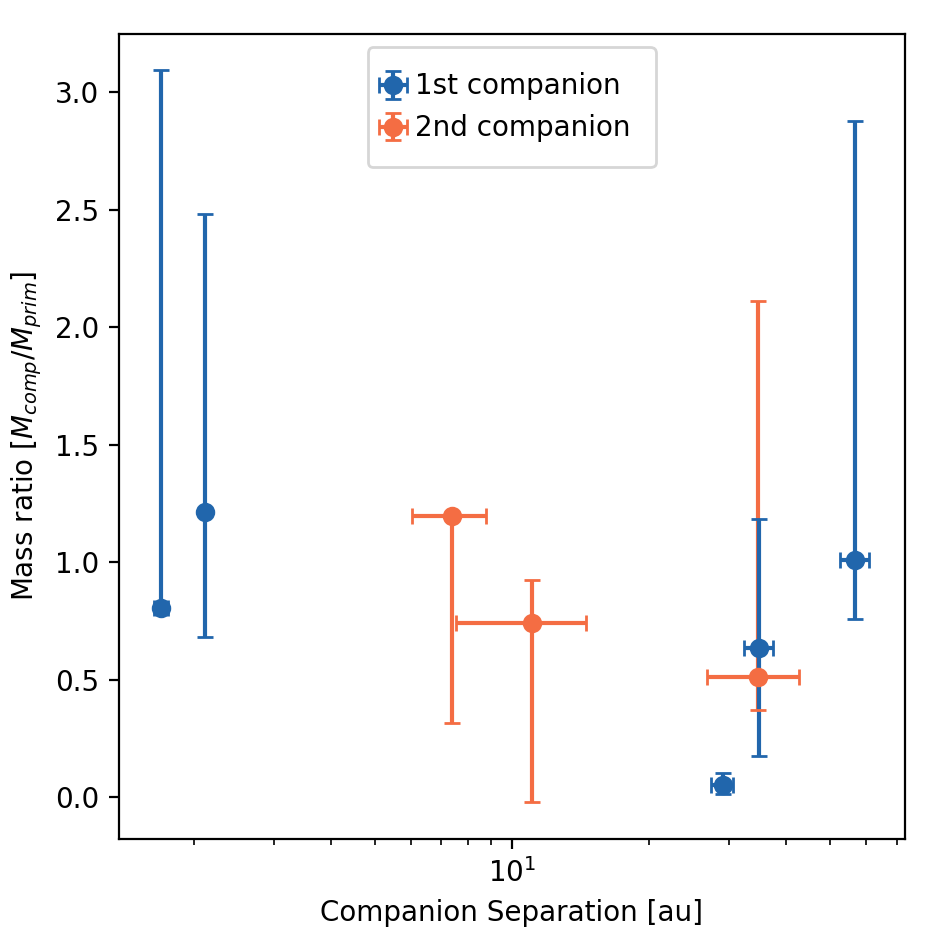}
\caption{Mass ratio per companion separation (in au). In blue, the datapoint refers to the mass ratio calculated with the closest companion. In orange, the mass ratio have been calculated with the second-closest companion found with GRAVITY. }
\label{fig:massratio}
\end{figure}

In Fig. \ref{fig:massratio}, we display the resulting mass ratios ($M_\mathrm{comp}/M_\mathrm{prim}$) with the corresponding companion separation. Again, no clear correlation or significant trend is found between mass ratio and separation, mainly due to the large errorbars involved in the appreciation of the data. 
In general, the calculated mass ratios indicate massive systems, regardless of their separation. The only systems in M17 with mass ratio close to 1 have either a small separation of $\sim$2.3 au or large separation $\sim$120~au. Intermediate mass ratios, from $\sim0.05$ to $\sim0.75$ are found for separations ranges 1.7 to 30~au. With comparing to Fig. \ref{fig:deltamag_vs_sep}, we cannot certify a correlation between flux ratio and mass ratio, as some sources may be more extincted than others, depending on their position within the cluster. The extinction being an important parameter to be taken into account while retrieving the mass of the components from their spectra (See Sect. \ref{subsection:spectral_fitting}). 

An aside can be made concerning one of the sources for which VLT/XSHOOTER spectroscopic data are available: B111. Despite the high uncertainty in log($g$), \citet{Ramirez+2017} derived a spectroscopic mass of $42{\substack{+12 \\ -13}}$ \Msun, by using the stellar properties derived from best-fits FASTWIND parameters and the SED fitting. The mass of the primary derived in this study is $48\pm 1$\Msun, while the companion has a mass of $3\pm 2$\Msun. The overall mass of the system shows consistency with the spectroscopic mass derived by \citet{Ramirez+2017}.  

\subsection{Connection with star formation models}

We have obtained the first constraints on the multiplicity properties of young massive stars in the star-forming region of M17. Those stars are suspected to have already reached the zero-age main sequence (ZAMS). The system multiplicity will not be impacted by formation nor accretion mechanisms anymore. Only the effects from stellar and cluster evolution are going to have the most significant effects on the binary statistics. Here, we compare our results with predicted star formation scenarios. 
With 9 companions resolved by GRAVITY within 70~mas and a physical distance to M17 estimated at 1.7~kpc with {\it Gaia eDR3}, every star in our sample has at least one companion at physical distance less than 120~au. This coincides with \citet{Sana+2014} who found that 100\% of dwarfs in cluster or OB associations have a companion within 100~au. It seems clear that high-mass stars predominantly form in stellar systems/clusters, as opposed to in isolation. This raises questions about the origin of such a multiplicity and the formation of close binary stars {\it in situ}.

There is no consensus on the formation of high-mass binaries nor their characteristics at birth. To date, the most prominent theories leading to binary formation are stellar migration or capture, core accretion and fragmentation or disk fragmentation. Most massive star formation scenarios agree on the need for a dense and massive primordial cloud to overcome the radiative forces generated by the newborn star \citep{Zinnecker+2007}. High infall rates lead to massive and gravitationally unstable disks. Rapid accretion coupled with high angular momentum causes these disks to fragment \citep{Kratter+2006}, possibly producing low-mass companions at 100-1000~au \citep{Krumholz+2016}. While taking into account radiative heating, \citet{Krumholz2006} suggests that high temperatures stabilise the massive core, reducing the fragmentation and leading to small multiple systems, with high mass components. As a large volume to form such massive core fragment is needed, one would expect a correlation between system mass and separation.

As shown earlier (see Sect. \ref{section:Discussion}), the multiplicity properties across the main sequence is not uniform. Unlike the standard formation models resulting in the formation of single solar-type stars, the multiplicity rate is key to discriminate between massive star formation theories. Three main competing theories of massive star formation are under active scrutiny: stellar collisions and mergers, monolithic collapse in isolated cores and competitive accretion \citep{Zinnecker+2007}.

Stellar collisions could create a very small fraction of high-mass stars but requires extremely dense young clusters ($>10^{6}$ stars pc$^{-3}$, \citealt{Krumholz+2007,Zinnecker+2007}). With only several thousands sources located in the massive stellar cluster \citep{Broos+2007,Hoffmeister+2008}, stellar collisions and mergers cannot be a dominant formation pathway in M17 nor in most galactic clusters and OB associations, thus will be discarded in the following discussion. 

Competitive accretion, as developed by \citet{Bonnell+2005}, predicts that the forming star accretes material that was not initially gravitationally bound to the stellar seed (by contrast to the monolithic collapse). The latter competes for the mass available in the clump and grow by Bondi-Hoyle accretion \citep{Bonnell+2001,McKee+2003}. Under this framework, \citet{Bonnell+2005book,Bonnell+2005proceeding} predicts the formation of close binary systems through three-body capture. A massive primary star bounded with a lower mass star interacts with a massive wide companion. During close approaches, huge amounts of energy are exchanged resulting in the ejection of the lowest mass star, and a remaining central massive binary system. Accretion from a turbulent medium onto a binary can significantly tighten the system at the same time as it increases its mass ($a_\mathrm{bin}\propto M_\mathrm{bin}^{-2}$), meaning that a low-mass wide ($\sim$100~au, \citealt{Krumholz+2007}) binary is likely to turn into a high-mass close binary. This is driven by continuous accretion that hardens binaries while increasing the masses of the components \citep{Krumholz+2007}. The dependence of the separation on the system mass results in clusters, with high-mass stars located in the cluster centre with close massive companions. Gas accretion shrinks the orbits and the newborn system is composed of a high-mass close binary surrounded by a third low mass component at larger distances \citep{Bate+2002}. Yet, three-body interactions tend to preferentially keep more massive stars in binaries, thus favouring equal mass ratios \citep{Krumholz+2016}. In our sample, the formation of tight binaries (B189NE,B98) and the presence of high mass systems in the cluster core (the B189 complex) are in favour of competitive accretion. However, we do not observe massive twin binaries and we detect companions spanning a wide range of separations, from 1.7 to $\sim$120~au. As such, there is no preference for close massive systems: some massive stars have close companions (within 5~au) but other have wider companions (up to $\sim$120~au). In addition, no correlation between mass and separation is found. We observe both massive close companions as well as massive wide components. All of the stars in our sample are early-type, limiting our study to massive components, as opposed to the sample of \citet{GravityCollab+2018} in Orion probing a large variety of masses. A large spectrum of stars within the cluster is indeed expected by the competitive accretion scenario. The lack of observations of equal mass systems and the lack of relation between binary separation and system mass ($a_\mathrm{bin}\propto M_\mathrm{bin}^{-2}$) are in conflict with competitive accretion as the dominant formation channel.

In monolithic collapse models (extended version of standard low-mass star formation theory), \citet{McKee+2002,McKee+2003} theorise that a massive star (either individual or as a future system) originates from the collapse of a dense and massive core, that contains the mass reservoir available to form the star \citep{Krumholz+2007}. While collapsing, a protostellar disk is naturally formed around the core, with typical sizes of $~$1000~au. Those disks have strong spiral arms and are subject to fragmentation, per chance surviving and ending up as a companion star \citep{Krumholz+2007}. Such fragments tend to migrate inwards as the disk accretes. However, unlike the competitive accretion models, core accretion provides a framework in which protostars gain mass without tightening the binary separation \citep{Zinnecker+2007}.

Due to the high luminosity of the forming massive star, both scenarios agree on the need of high accretion rates to prevent the radiation pressure and the ionisation to restrain mass infall onto the central star \citep{McKee+2003}. High infall rates lead to the formation of gravitationally unstable circumstellar disks with polar collimated outflows, overcoming the radiation pressure and allowing the growth of the central star. This disks are eventually prone to fragmentation \citep{Kratter+2006} and possibly producing low-mass companions at 100-1000~au \citep{Krumholz+2016}.
With suspecting a scaled-up version for massive stars where the latter grow by disk accretion, \citet{Oliva+2020} studied the formation of companions from spectroscopic regimes to 2000~au, based on simulations. A growing accretion zone is formed around the massive newborn star, then forms spiral arms and shortly afterwards, fragments. These fragments are highly dynamic structures that interact with other fragments, the disk itself and spiral arms. Continuously interacting with their environment, most of the fragments are likely destroyed.  
However, a few of them grow in mass over time, form second Larson cores and yield companions in the middle or outer regions of the disk. The separation of the newborn companions should be comparable to disk sizes, which are typically 100s to 1000s of au, at the latest stages of the star formation process. Such protostars are prone to inward or outward migration in the long term \citep{Meyer+2018}. In the context of disk fragmentation theories, such an inward migration process may be driven by the interaction with the remnant of the accretion disk or with protostellar bodies of the cluster.

The wide range of properties obtained among the M17 population in terms of separations and masses does not allow us to identify a single (or mostly dominant) formation mechanism. The results agree with different formation scenarios with respect to different separation ranges. While tight massive binaries could be explained with failed (migration induced) mergers, by accretion onto protobinaries or induced by the evolution of wider systems, massive wide binaries could be the product of fragmentation and disk-assisted capture \citep{Zinnecker+2007}.

Nonetheless, it is worth mentioning that given the young age of M17 ($<$1~Myr), some dynamic processes may still be undergoing, and that the final stellar content and structure of M17 is not set yet.
In both accretion scenarios, competitive or core accretion, wide binaries are initially formed at large distances and eventually harden over time. Observing companions close to the primary star could also indicate that protostellar cores that were formed at the edges of the accretion disk, migrated over time, as suggested by \citet{Oliva+2020} and \citet{Ramirez+2021}. This hardening process that may occur on timescales of the order of 2~Myr \citep{Ramirez+2021} might be driven by the interaction with the remnant of the accretion disk in the context of disk fragmentation theories, or with other protostellar bodies present in the cluster. Radial velocity examinations in M17 suggested that there is a lack for short-period binaries (close companions) in the cluster, to which \citet{Sana+2017} proposed that the migration scenario, as the main process to shrink binaries periods, might harmonise observational results in M17 and multiplicity properties observed in few million year-old OB-star populations. In fact, being about $\sim$1~Myr-old, the migration process inside M17 is likely underway: the wide range of separations and mass ratios might illustrate that the stellar population did not reach the expected time to draw the final structures. A combination of mechanisms is at play, and further investigations among the youngest objects is crucial. If the migration scenario is correct, we indeed expect that most of the objects have at least one detectable companion within the expected size of the accretion disk. Non-detection of companions within $\sim$200~au would invalidate the proposed scenario. 

We note that not all objects need to undergo migration. The spectroscopic binary fraction reaches $\sim$50\% for B stars to $\sim$70\% for O stars, meaning that there also exist long-period massive binaries. Indeed, the period distribution of massive binaries at an age of a few Myr follows a well defined power low ranging from about 1~d to 10~years at least \citep{Fryer+2012,Sana+2012,Sana+2014,Kobulnicky+2014,Almeida+2015,Banyard+2021,villasenor2021}. A migration affecting about half to two thirds of all OB stars is therefore probably sufficient to explain the wide variety of separation observed. 

nt medium onto a binary can significantly tighten the system at the same time as it increases its mass ($R_\mathrm{bin}\propto M_\mathrm{bin}^{-2}$), meaning that a low-mass and wide ($\sim$100~au,\citet{Krumholz+2007}) binary is likely to turn into a high-mass close binary. This is driven by continuous accretion that hardens binaries while increasing the masses of the components \citep{Krumholz+2007}. The dependence of the separation on the system mass results in clusters, with high-mass stars located in the cluster center with close massive companions. Under this framework, \citet{Bonnell+2005book,Bonnell+2005proceeding} predicts the formation of close binary systems through three-body capture. A massive primary star bounded with a lower mass star interacts with a massive wide companion. During close approaches, huge amounts of energy are exchanged resulting in the ejection of the lowest mass star, and a remaining central massive binary system. Gas accretion shrinks the orbits and the newborn system is composed of a high-mass close binary surrounded by a third low mass component at larger distances \citep{Bate+2002}. Yet, three-body interactions tend to preferentially keep more massive stars in binaries, thus favoring equal mass ratios \citep{Krumholz+2016}. With forming close associations with fairly high eccentricities in dense star-forming regions ($10^{6}$ stars pc$^{-3}$, \citealt{Krumholz+2007}), periastron separations can dangerously close the size of the stellar radii. Those systems could eventually merge and, under the assumption of a successful process, leave a higher mass single star \citep{Bonnell+2005}.

\section{Conclusions}
\label{section:Conclusion}

In our quest to understand the origin of close massive stars, we performed VLTI/GRAVITY observations of six young O-stars in NGC6618 cluster, inside M17. The high angular resolution reached by the VLTI allowed us to fill the gap of the unexplored range 1$-$100~mas, between spectroscopic campaigns and imaging techniques. As the sources are all O-type objects of luminosity class V, they have already reached the main sequence. We used PMOIRED to model the visibility amplitudes and closure phases of each object in order to derive their multiplicity parameters. We identified the presence of binaries and triples, with companions orbiting within $\sim$120~au.

\begin{itemize}
    \item All of the objects in M17 appear to be multiple systems. Hence, we measure a multiplicity fraction of 100\%, with MF$>$94\% at the 68\%-confidence interval. 
    \item We detect a total of 9 interferometric companions: we resolve two companions around B189SW, B0 and B260, with magnitude differences up to 4 and 1 companion around B189NE, B98 and B111 at relatively close distances from their host stars (within $\sim$30~au).
    \item Combined with the companions reported in the literature, based on spectroscopic surveys, a total of 14 companions is found. As such, we derive a companion fraction of $2.3\pm0.6$ for our complete sample. 
    \item With comparing our data to atmosphere models and evolutionary tracks, we derive some stellar parameters, such as log($T$), the age and the mass of each star. A large range of masses have been derived, from 3 to 50~\Msun. 
     \item The multiplicity and companion fraction derived in this region of M17, together with the mass estimations, further confirm that the first two quantities quoted rise with primary mass. Despite its young age, M17 shows consistency with previous studies \citep{Sana+2014,GravityCollab+2018}.
    \item Finally, our observational results are compared to the expected properties of star formation predictions. Neither core accretion nor competitive accretion can fully reproduce our observations. A combination of processes may better explain the variety of separations and masses found in M17, among the population of young O-type stars. The picture will better be completed with the observation of the youngest stars in the cluster, i.e most embedded sources, the ones expected to better represent the outcome of star formation for high-mass stars. 
\end{itemize}

\begin{acknowledgements}
This research has received funding from the European Research Council (ERC) under the European Union’s Horizon 2020 research and innovation programme (grant agreement number 772225: MULTIPLES). Based on observations collected at the European Southern Observatory under ESO programme 0101.C-0305(A) and 0101.C-0305(B).
This work has made use of data from the European Space Agency (ESA) mission \emph{Gaia} (\url{https://www.cosmos.esa.int/gaia}). This research has made use of the Jean-Marie Mariotti Center \texttt{SearchCal}\footnote{Available at http://www.jmmc.fr/searchcal}, \texttt{OIFits Explorer}\footnote{Available at http://www.jmmc.fr/oifitsexplorer} service, \texttt{LITpro}\footnote{\texttt{LITpro}software available at http://www.jmmc.fr/litpro} and \texttt{OiDB} service \footnote{Available at http://oidb.jmmc.fr } service co-developped by CRAL, IPAG and LAGRANGE. This research has made use of the SIMBAD database, operated at CDS, Strasbourg, France. This research made use of Astropy,\footnote{http://www.astropy.org} a community-developed core Python package for Astronomy \citep{astropy:2013, astropy:2018} and of PMOIRED (\url{https://github.com/amerand/PMOIRED}), developed by Antoine Mérand. We used the internet-based NASA Astrophysics Data System for bibliographic purposes.
\end{acknowledgements}

\bibliographystyle{aa}
\bibliography{references}

\newpage
\begin{appendix}

\section{GRAVITY observations and best-fit model for each source}
We present single epoch GRAVITY observations of the six young O stars in the M17 region and the two foreground stars (initially in the sample). The visibilities and closure phases are plotted as a function of the interferometric baselines. Some of the sources (B189NE, B111, B260) show remaining signal in the residuals while modeling the visibilities. Although the residuals are mainly centered around 0, this shape might be the result of instrumental biases or the lack of geometric complexity while performing the fit. Multiple tests including more complex models with asymmetric or slanted disks (instead of a regular fully resolved component) have been applied in the attempt to correct these residuals. None of these trials were conclusive, in particular because they did not improve the quality of Chi2. With getting more than one snapshot per target with an enhanced (u,v) coverage we could perform image reconstruction, that would help in understanding whether these residuals are instrumental or involve a more complex interferometric signal.

For B189SW, we also display the model image showing the disk around the second companion at $\sim$120~au. The best-fit model for each source is the one highlighted in bold in Table \ref{table:best_fit_tot} except for B189SW, the overlaid fit is the one presented in Table \ref{tab:B189SW_fit}.
\label{appendix:GRAVITY_DATA}

\begin{figure}[!htb]
    \subfloat[TYC-6265-1977-1]{{\includegraphics[width=.48\textwidth]{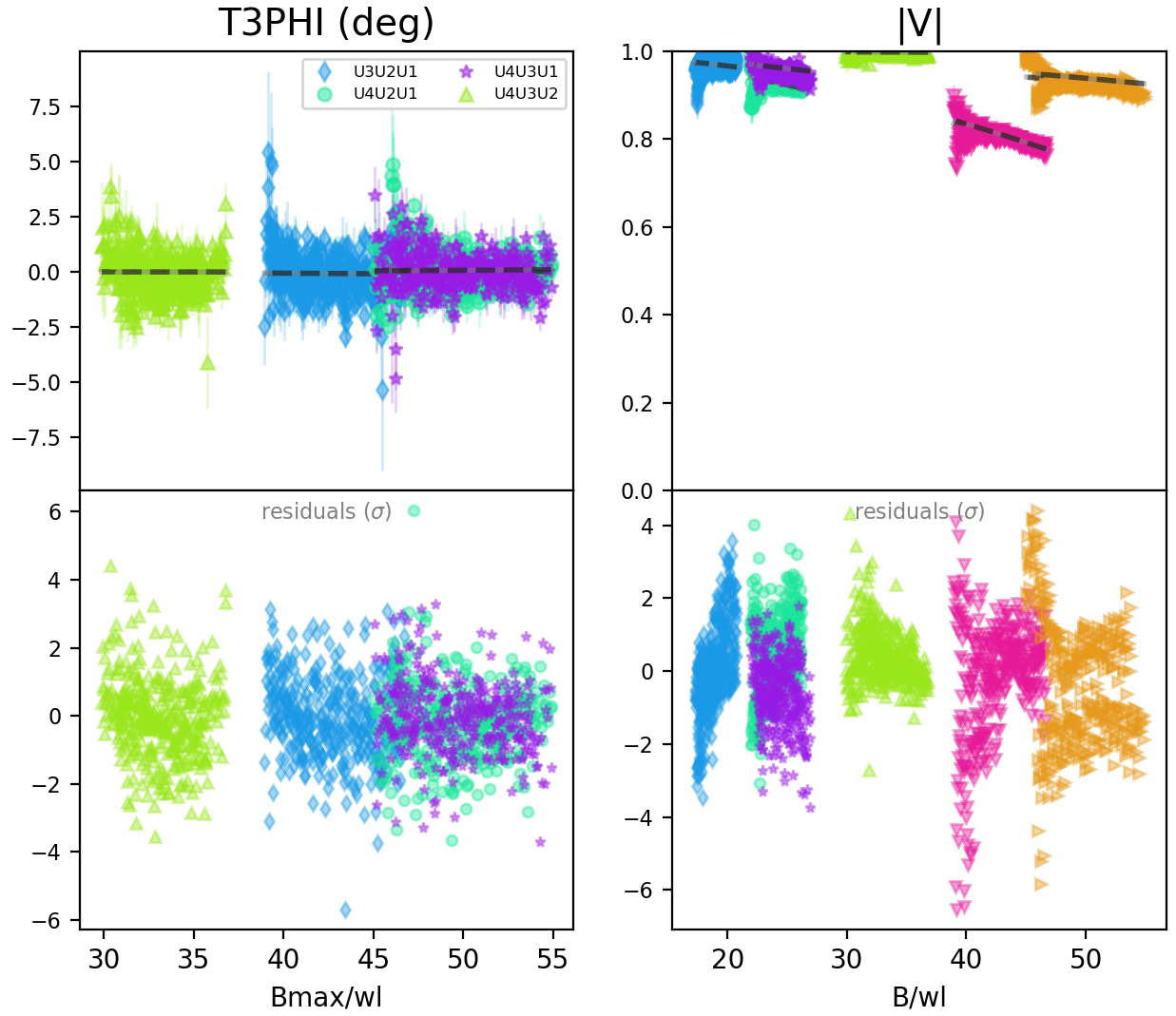}}} \\
    \subfloat[NGC6618-B293]{{\includegraphics[width=.48\textwidth]{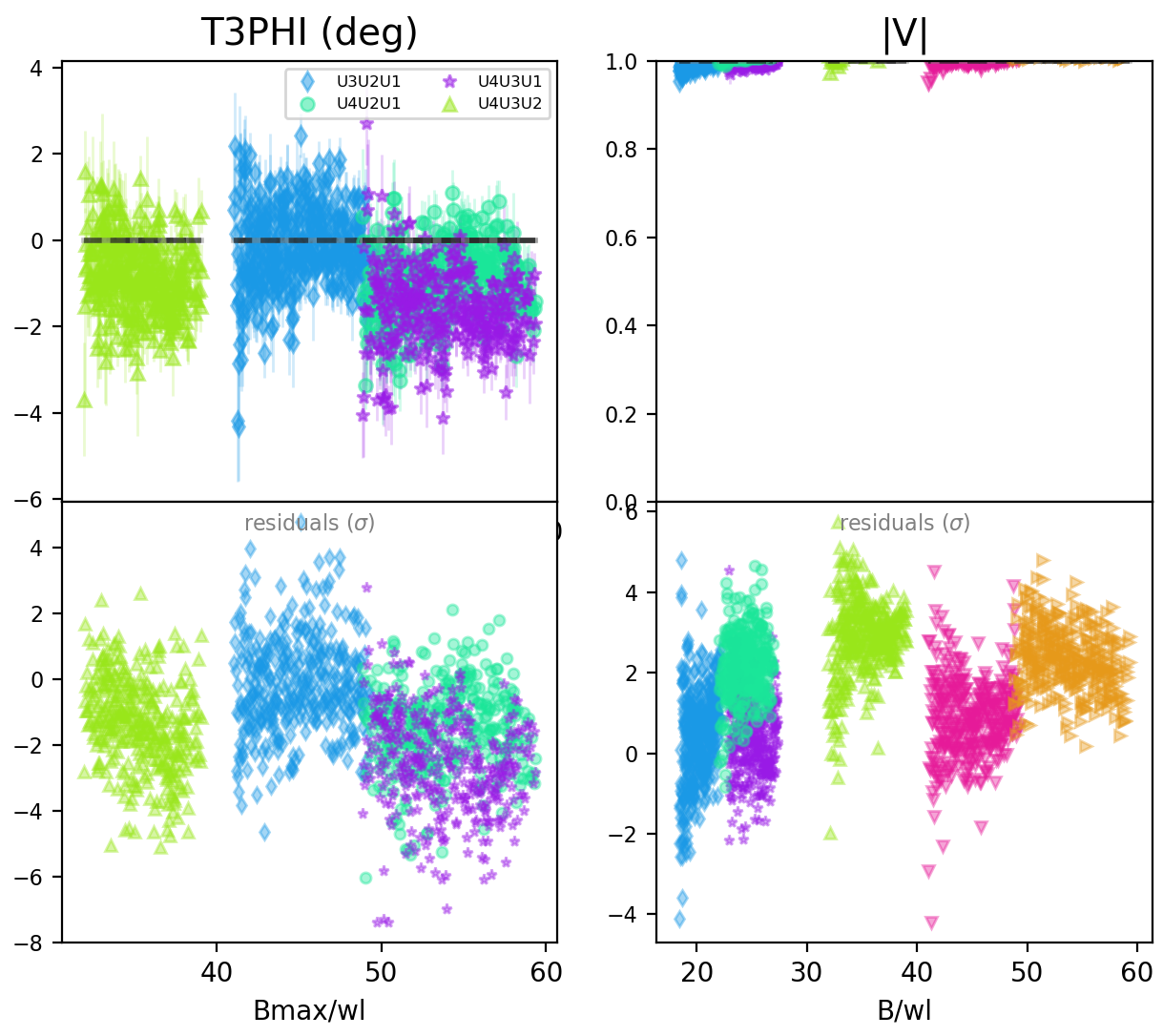}}}
	\caption{Summary plot of the GRAVITY observations for the two foreground stars overplotted with the best-fit model in dark grey lines. The fit is performed over a restricted wavelength range from 2.05 to 2.45~\si{\micro\meter}. The colours code the baseline for visibilities, and the triangle for the closure phases.}
	\label{fit:mycaption}
\end{figure}
\FloatBarrier
\begin{figure*}[!h]
	\subfloat[NGC6618-B189SW]{\includegraphics[scale=0.20]{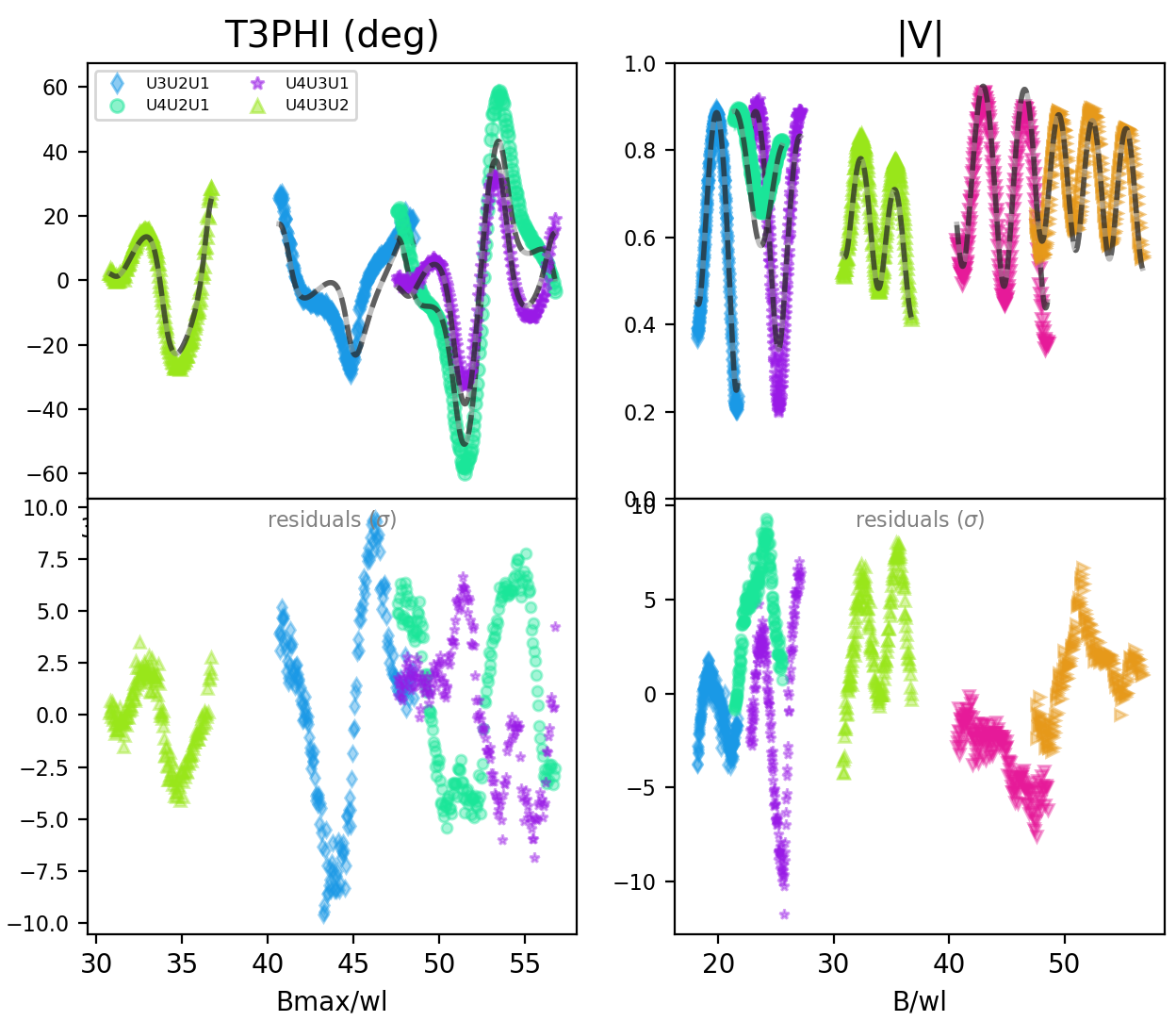}}
	\subfloat[NGC6618-B189NE]{\includegraphics[scale=0.20]{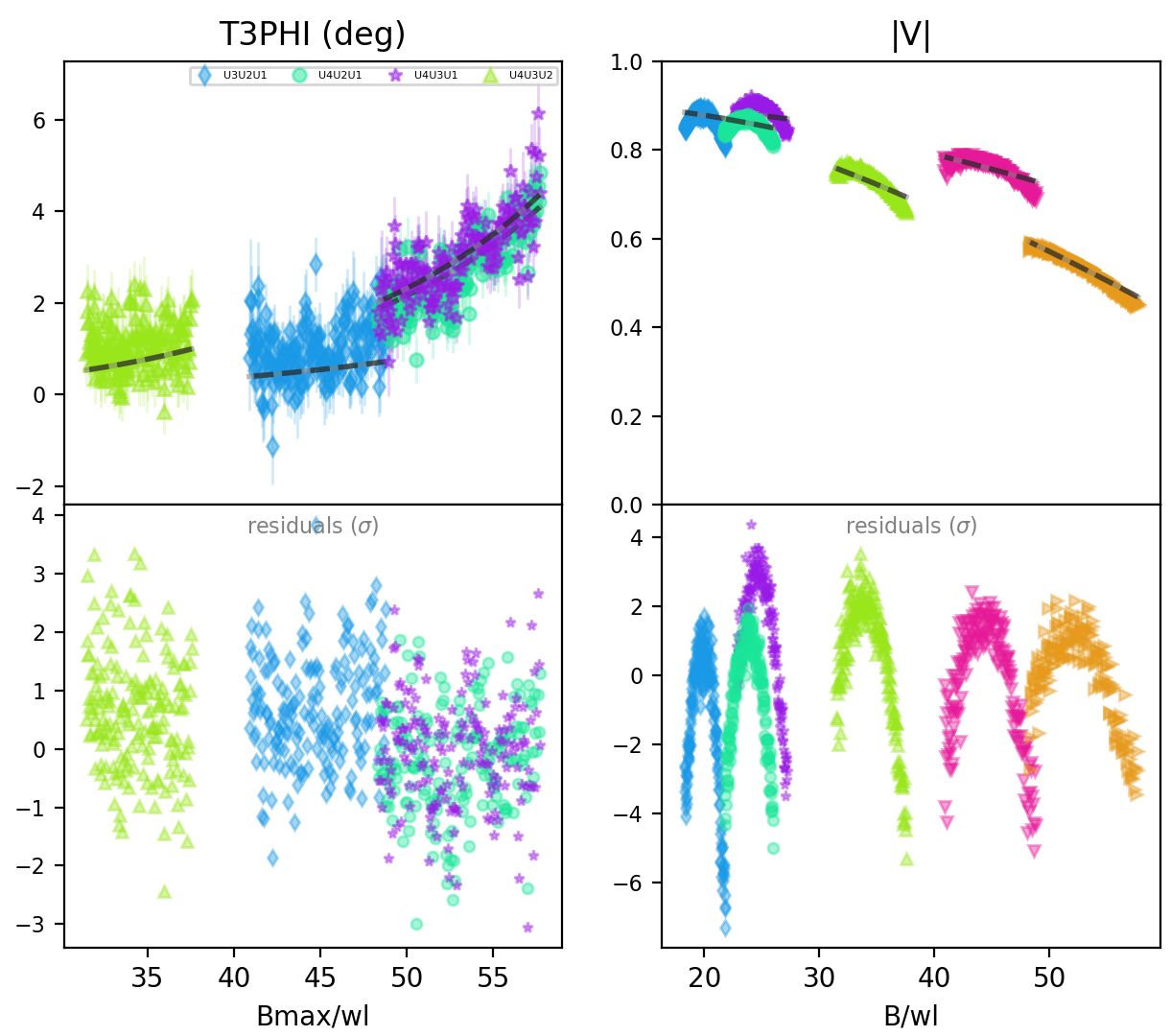}}\\
	\subfloat[NGC6618-B111]{\includegraphics[scale=0.20]{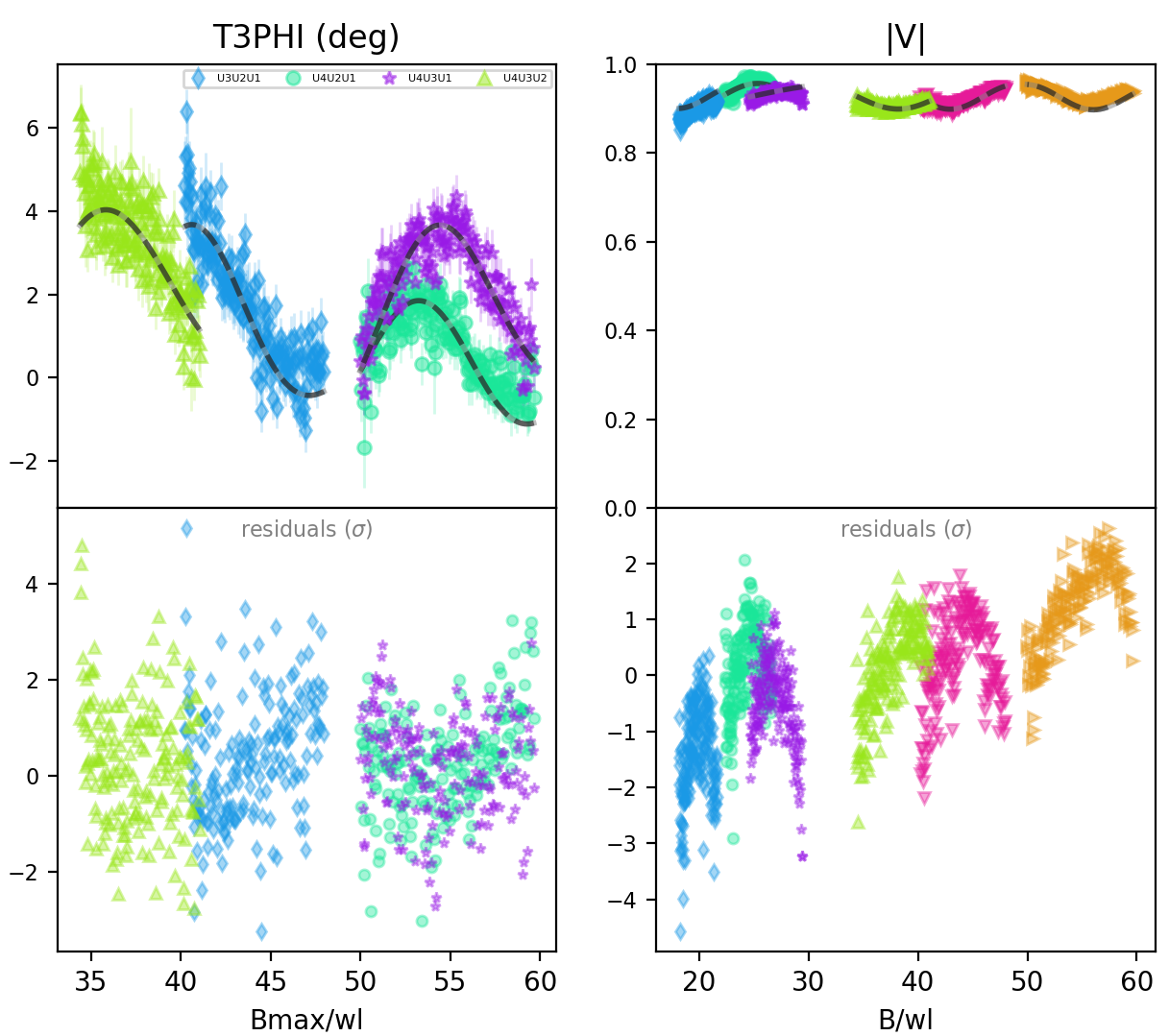}}
	\subfloat[NGC6618-B98]{\includegraphics[scale=0.20]{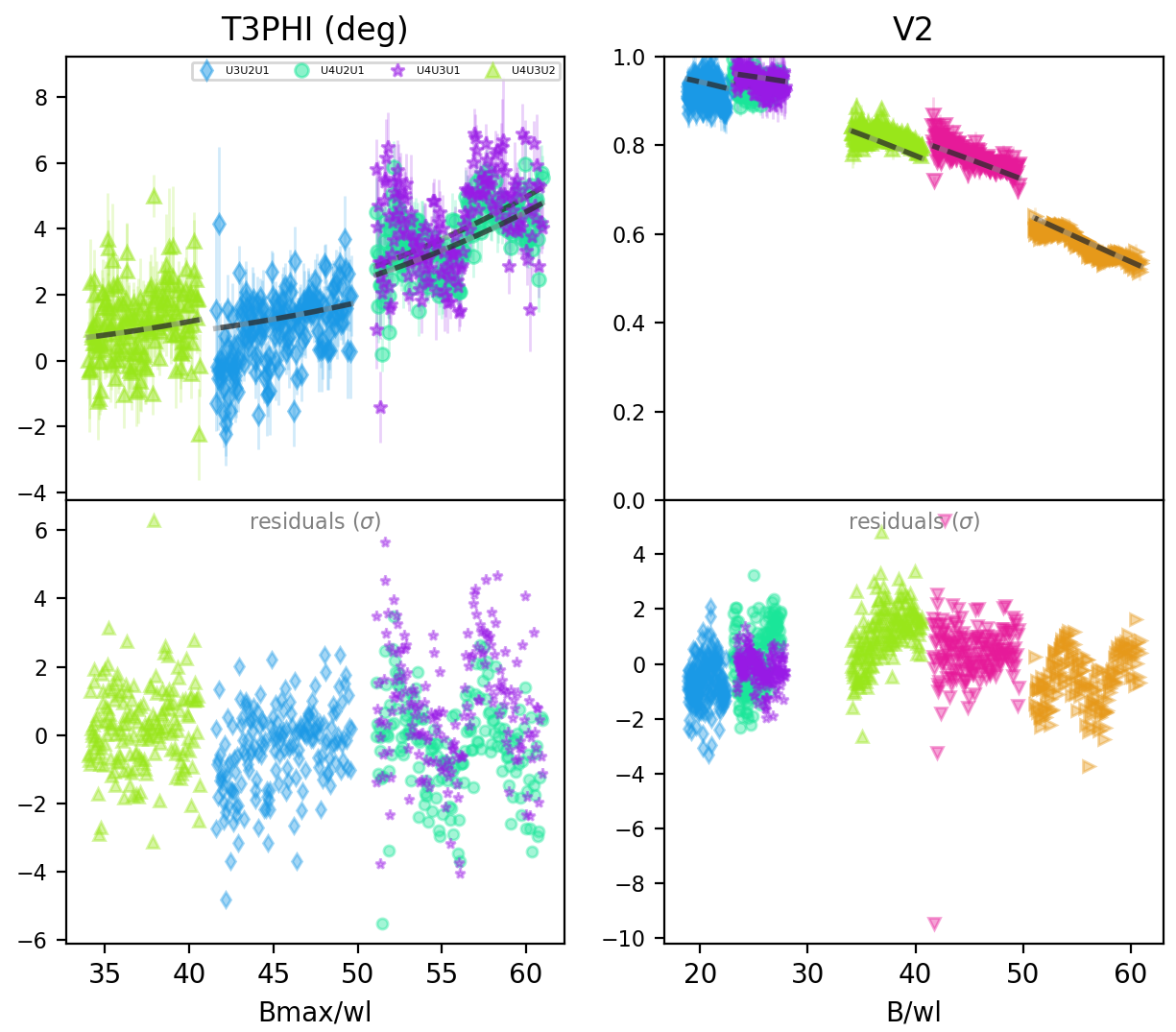}}\\
	\subfloat[NGC6618-B0]{\includegraphics[scale=0.20]{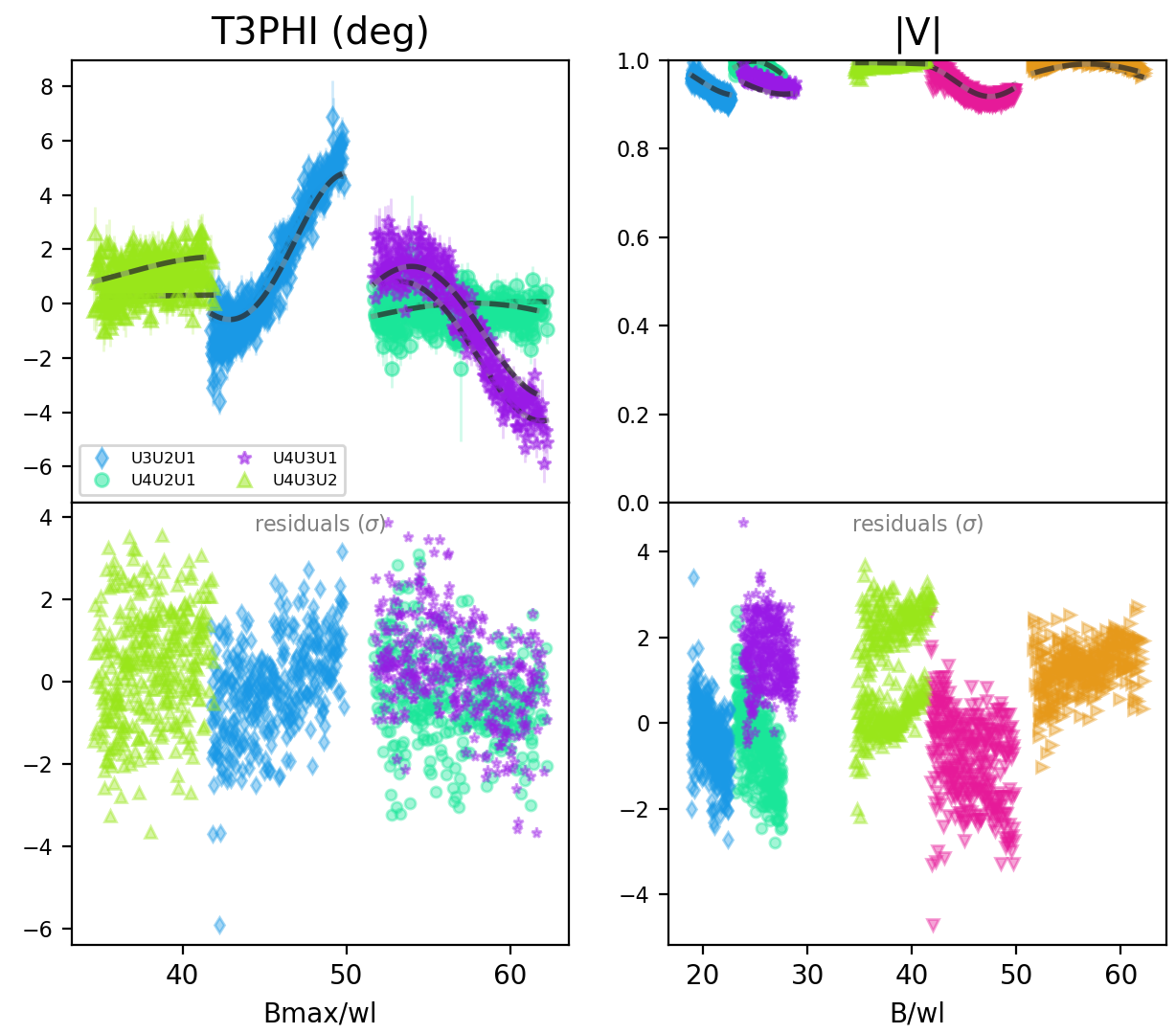}}
	\subfloat[NGC6618-B260]{\includegraphics[scale=0.20]{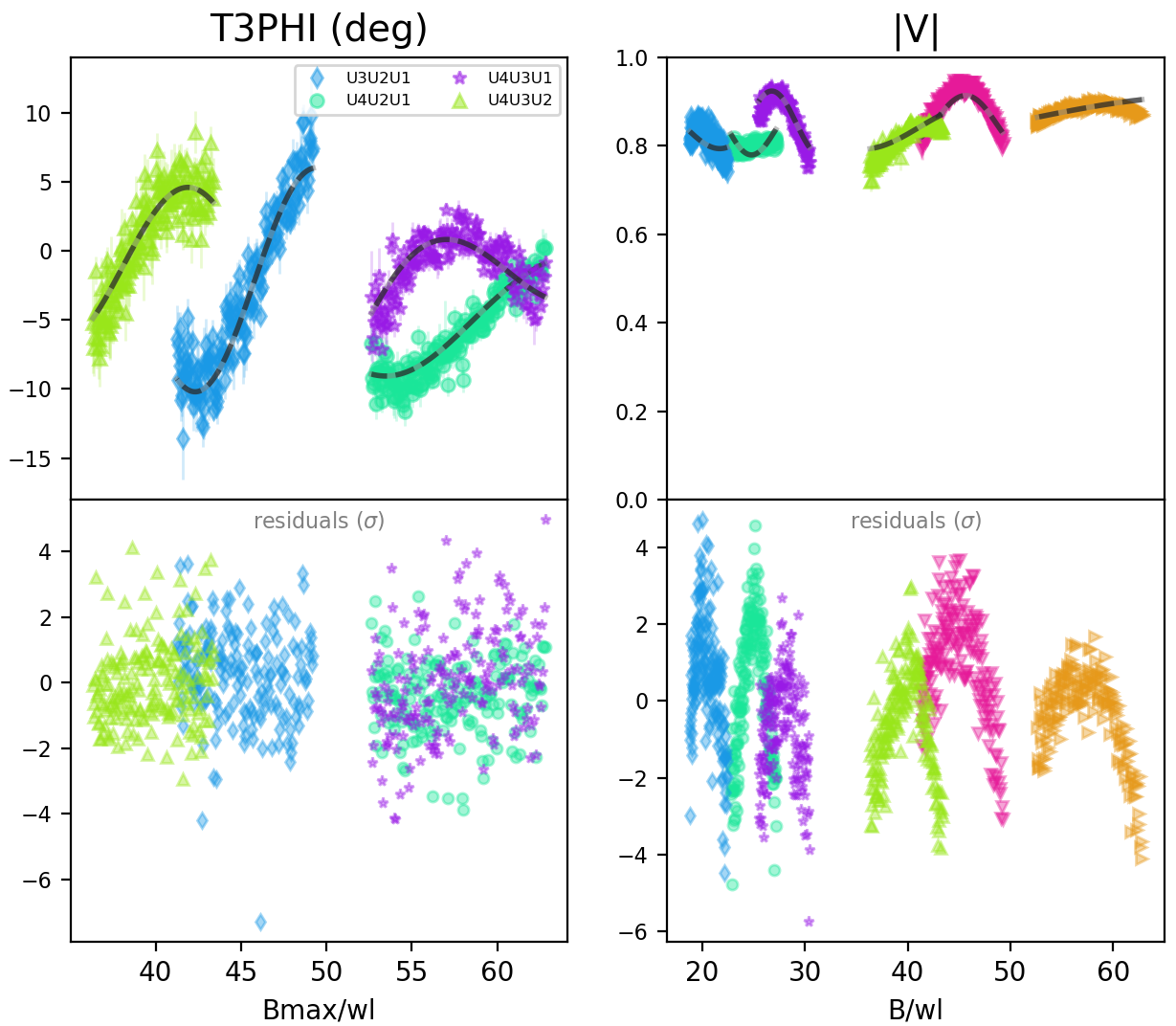}}
	\caption{Summary plot of the GRAVITY observations for the six young O stars in M17 overplotted with the best-fit model in dark grey lines. The fit is performed over a restricted wavelength range from 2.05 to 2.45~\si{\micro\meter}. The colours code the baseline for visibilities, and the triangle for the closure phases. }
	\end{figure*}

\FloatBarrier
\section{Best-fit parameters and $\chi^{2}_{r}$ progression}
\begin{figure}[!h]
    \centering
    \includegraphics[scale=0.3]{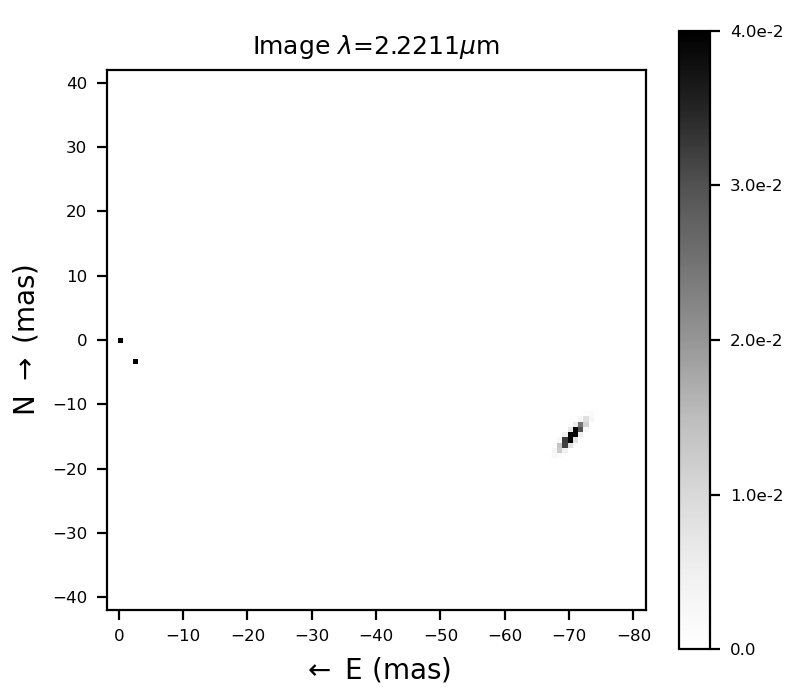}
    \caption{Illustration of the three components involved in the B189SW system, based on the best fitting geometric model to the GRAVITY data presented in Table \ref{tab:B189SW_fit}.}
    \label{fig:B189_SW}
\end{figure}

We present here the results of the fitting process for GRAVITY data of the M17 sources and the two foreground stars TYC6265 and B293 (see Table \ref{table:best_fit_tot}). After testing the single star model on each source, we further investigate the morphology of the source with the goal to determine the presence of one or multiple companion(s). The best binary astrometry is found with searching for the global minimum in a 2D grid of positions (as explained in Sect. \ref{section:DataAnalysis}). We iteratively repeat the process while searching for a putative second companion. Our algorithm also allows the attribution of a fraction of the total flux to a fully resolved emission component (hereafter: Fully res. component). 
The model-fitting result retained for B189SW, involving a more complex geometry, is presented in Table \ref{tab:B189SW_fit} and figure \ref{fig:B189_SW}.
\label{appendix:Fit_parameters}

\FloatBarrier
\begin{table}[!h]
    \centering
    \begin{tabular}{lccc}
    \hline
    \hline \\ [-1ex]
         \multicolumn{4}{c}{First companion, A} \\[1ex]
    \hline \\[-1ex]
         Position to East & $\Delta x_{A}$ & $(mas)$ & $-70.42\pm0.04$ \\
         Position to North & $\Delta y_{A}$ & $(mas)$ & $-14.80\pm0.05$ \\
         Flux ratio          & $F_{A}/F_{cs}$  &          & $0.44\pm0.04$\\
         Spectral index     & $\alpha$      &           & $-2.4\pm0.5$ \\
         Gaussian FWHM size & $\Sigma_{A}$ & $(mas)$ & $3.97\pm0.41$ \\
         Gaussian PA        & $\theta_{A}$ & $(^\circ)$ & $42.8\pm3.6$ \\
         Gaussian inclination   & $i_{A}$ & $(^\circ)$ & $77.7\pm5.9$ \\[1ex]
    \hline \\ [-1ex]
         \multicolumn{4}{c}{Second companion, B}\\[1ex]
    \hline \\ [-1ex]
         Position to East & $\Delta x_{B}$ & $(mas)$ & $-2.52\pm0.15$ \\
         Position to North & $\Delta y_{B}$ & $(mas)$ & $-3.61\pm0.20$ \\
         Flux ratio          & $F_{B}/F_{cs}$   &           & $0.09\pm0.02$\\
         Spectral index     & $\alpha$      &           & $-4.0\pm1.3$ \\[1ex]
     \hline \\ [-1ex]
        Reduced $\chi^{2}$ & \bm{$\chi_{r}^{2}$}  & & \textbf{14.2} \\ [1ex]
    \hline 
         
    \end{tabular}
    \caption{Model-fitting result for B189SW, including two unresolved components and an inclined Gaussian. The central component, set as an unresolved punct at the coordinates $(0,0)$, is used as the phase and flux reference ($F_{cs}$). }
    \label{tab:B189SW_fit}
\end{table}

\FloatBarrier

\FloatBarrier
\begin{sidewaystable*}
\centering
\caption{Models and $\chi_{r}^{2}$ progression for each source.}
\resizebox{\textwidth}{!}{%
\begin{tabular}{@{\extracolsep{5pt}}cccccccccccc@{}}
    \hline
    \hline 
    \\ [-1.5ex]
    & & & \multicolumn{3}{c}{Binary} & \multicolumn{3}{c}{Triple} & & \\
    \cline{4-6}  \cline{7-9}
    \\ [-1.5ex]
    Source & \#Fit & Unresolved &  $\Delta x_{2}$ & $\Delta y_{2}$ & $f_{r1}$ & 
                                    $\Delta x_{3}$ & $\Delta y_{3}$ & $f_{r2}$ &
                                  Fully res. & $N_{D.o.F}$ & $\chi_{r}^{2}$ \\
           &       & Component 1 &  & & &
                                     & & &
                                     Component & \\
    \hline \\ [-1.5ex]
             &       &  Uniform Disk & Position & Position & &
                                    Position & Position & &
                                    Flux & \\
            &       &  $(mas)$  & $(mas)$ to East  & $(mas)$ to North & $F_\mathrm{2}/F_\mathrm{1}$ &
                                  $(mas)$ to East  & $(mas)$ to North & $F_\mathrm{3}/F_\mathrm{1}$ &
                                  & \\
    \hline \\ [-1.5ex]
    \multirow{3}{*}{B189SW\tablefootmark{a}} & 0 & $2.62\pm0.31$ & - & - & - & - & - & - & - & 1812 & 392.2 \\
                                     & 1 & 0.2 & $-70.43\pm0.08$ & $-14.81\pm0.11$ & $0.37\pm0.04$ & 
                                     - & - & - & - & 1809 & 114.3 \\
                                     & 2 & 0.2 & $-70.34\pm0.05$ & $-14.86\pm0.07$ & $0.32\pm0.03$ & 
                                     $-2.64\pm0.28$ &$-3.36\pm0.48$ & $0.12\pm0.01$ & - & 1806 & 48.7 \\
    
    \hline \\ [-1.5ex]
     \multirow{3}{*}{B189NE} & 0 & $2.58\pm0.21$ & - & - & - & - & - & - & - & 1811 & 36.3 \\
                                     & 1 & 0.2 & $1.38\pm0.08$ & $0.08\pm0.20$ & $0.82\pm0.03$ & 
                                     - & - & - & - & 1809 & 25.7 \\
                                     & \textbf{2} & \textbf{0.2} & \bm{$1.26\pm0.02$} & \bm{$0.08\pm0.03$} & \bm{$0.75\pm0.01$} & 
                                     - & - & - & \bm{$0.15\pm0.01$} & \textbf{1808} & \textbf{2.7} \\
    
    \hline \\ [-1.5ex]
    \multirow{3}{*}{B111} & 0 & $1.25\pm0.21$ & - & - & - & - & - & - & - & 1811 & 23.0 \\
                                     & 1 & 0.2 & $-10.45\pm0.24$ & $-13.63\pm0.27$ & $0.05\pm0.01$ & 
                                     - & - & - & - & 1809 & 7.8 \\
                                     & \textbf{2} & \textbf{0.2} & \bm{$-10.35\pm0.05$} & \bm{$-13.81\pm0.06$} & \bm{$0.03\pm0.01$} & 
                                     - & - & - & \bm{$0.05\pm0.01$} & \textbf{1808} & \textbf{1.3} \\
    \hline \\ [-1.5ex]
    \multirow{2}{*}{B98} & 0 & $1.61\pm0.07$ & - & - & - & - & - & - & - & 1811 & 11.1 \\
                                     & \textbf{1} & \textbf{0.2} & \bm{$0.95\pm0.03$} & \bm{$0.35\pm0.04$} & \bm{$0.31\pm0.03$} & 
                                     - & - & - & - & \textbf{1809} & \textbf{1.7} \\
    \hline \\ [-1.5ex]
    \multirow{3}{*}{B0} & 0 & $0.68\pm0.16$ & - & - & - & - & - & - & - & 3618 & 15.5 \\
                                     & 1 & 0.2 & $-1.37\pm0.05$ & $20.72\pm0.07$ & $0.04\pm0.01$ & 
                                     - & - & - & - & 3615 & 1.7 \\
                                     & \textbf{2} & \textbf{0.2} & \bm{$-1.39\pm0.04$} & \bm{$20.80\pm0.06$} & \bm{$0.04\pm0.01$} & 
                                     \bm{$5.7\pm0.09$} & \bm{$-3.02\pm0.18$} & \bm{$0.010\pm0.005$} & - & \textbf{3618} & \textbf{1.6} \\
    \hline \\ [-1.5ex]
    \multirow{4}{*}{B260} & 0 & $1.53\pm0.24$ & - & - & - & - & - & - & - & 1812 & 79.7 \\
                                     & 1 & 0.2 & $14.11\pm0.38$ & $-29.30\pm0.26$ & $0.11\pm0.01$ & 
                                     - & - & - & - & 1809 & 25.3 \\
                                     & 2 & 0.2 & $14.25\pm0.19$ & $-29.41\pm0.25$ & $0.09\pm0.01$ & 
                                     $-11.61\pm0.92$ & $18.21\pm0.91$ & $0.05\pm0.01$ & - & 1806 & 9.9 \\
                                     & \textbf{3} & \textbf{0.2} & \bm{$14.09\pm0.06$} & \bm{$-29.35\pm0.08$} & \bm{$0.07\pm0.01$} & 
                                     \bm{$-10.68\pm0.19$} & \bm{$17.69\pm0.24$} & \bm{$0.020\pm0.005$} & \bm{$0.09\pm0.01$} & \textbf{1805} & \textbf{2.3} \\
   \hline
    \hline \\ [-1.5ex]
    \multirow{3}{*}{TYC-6265} & 0 & $1.42\pm0.19$ & - & - & - & - & - & - & - & 3620 & 20.7 \\
                                     & \textbf{1} & \textbf{0.2} & \bm{$-0.48\pm0.01$} & \bm{$1.13\pm0.01$} & \bm{$1.03\pm0.02$} & 
                                     - & - & - & - & \textbf{3618} & \textbf{1.6} \\
    \hline \\ [-1.5ex]
    B293 & \textbf{0} & \bm{$0.0{\substack{+3.1 \\ -0.0}}$} & - & - & - & - & - & - & - & \textbf{3621} & \textbf{4.3} \\
    \hline
\end{tabular}}
\tablefoot{
\tablefoottext{a}{The first unresolved component is set at position (0,0) by default and has a flux of 1. The fit \#0 marks the start of our fitting process and refers to a single-star model. The second and third components are unresolved puncts with a uniform disk diameter of 0.2~mas ensuring they are unresolved at VLTI baselines. The free parameters are their position ($\Delta x$, $\Delta y$) and the relative flux $f_{r}$. The best-fit model is highlighted in bold. A more detailed table presenting the best-fit model retained for B189SW can be found in Table \ref{tab:B189SW_fit}}}
\label{table:best_fit_tot}
\end{sidewaystable*}

\section{Master Table}
\label{appendix:Master_Table}
We present an overview of the six observed stars in the M17 region. We display the results of the best fits and the mass estimations.
\begin{sidewaystable*}
    \caption{Overview of the six young-O stars observed in M17.}
    \centering
    \small
    \begin{tabular}{cc  c  c  c  c  c  c  c  c  c  c  c}
    \hline
    \hline
    \\
     & & & & & \multicolumn{4}{c}{Interferometry} & \multicolumn{4}{c}{Mass Modelisation} \\ 
     \cline{6-8}\cline{10-13}
     \\
    Object & CEN ID & Component & Optical & $m_{K}$ & Separation & Phys. Separation & Flux ratio & $\chi^{2}$ & Mass & log($T$) & log(Lum) & $\chi^{2}$ \\
    & & & Spec. type & & (mas) & (au) & & & (\Msun) & & \\
    \hline
    \\
    
    NGC6618-B189SW & CEN 1a & 1 & O4V\tablefootmark{a} & 6.9\tablefootmark{b} &  &  &  &  & $28{\substack{+0 \\ -13}}$ & $4.41{\substack{+0.00 \\ -0.38}}$ & $0.73{\substack{+0.01 \\ -0.05}}$ & 17.6 \\
     \\
     &  & 2 &  & $7.79\pm0.23$ & $71.95\pm2.49$ & $120.88\pm4.18$ & $0.44\pm0.04$ & 14.2 & 
      - & - & - & -  \\
     \\
     &  & 3 &  & $9.51\pm0.55$ & $4.40\pm0.82$ & $7.40\pm1.37$ & $0.09\pm0.02$ & 14.2 & $33{\substack{+1 \\ -18}}$ & $4.59{\substack{+0.01 \\ -0.44}}$ &  $0.73{\substack{+0.02 \\ -0.08}}$ & 17.7 \\
     \\
    
    NGC6618-B189NE & CEN 1b & 1 & O4V\tablefootmark{a} & 6.9\tablefootmark{b} &  &  &  &  & $19{\substack{+17 \\ -5}}$ & $4.16{\substack{+0.26 \\ -0.13}}$ & $0.71{\substack{+0.04 \\ -0.03}}$ & 25.3 \\
    \\
     &  & 2 &  & $7.21\pm0.03$ & $1.26\pm0.02$ & $2.12\pm0.03$ & $0.75\pm0.01$ & 2.7 & $23{\substack{+13 \\ -9}}$ & $4.12{\substack{+0.03 \\ -0.09}}$ & $0.72{\substack{+0.03 \\ -0.04}}$ & 26 \\
     \\

     NGC6618-B111 & CEN 2 & 1 & O4.5V\tablefootmark{c} & 7.5 &  &  &  &  & $49.1{\substack{+0.3 \\ -0.3}}$ & $4.65{\substack{+0.01 \\ -0.0}}$ & $0.75{\substack{+0.0 \\ -0.0}}$ & 124.6 \\
     \\
     &  & 2 &  & $11.30\pm0.83$ & $17.26\pm0.97$ & $29.00\pm1.64$ & $0.03\pm0.01$ & 1.3 & $2.5{\substack{+2.5 \\ -2}}$ & $3.68{\substack{+0.3 \\ -0.1}}$ & $1.11{\substack{+1.6 \\ -0.3}}$ & 91 \\
    \\

     NGC6618-B98 & CEN 3 & 1 & O9V\tablefootmark{a} & 7.7 &  &  &  &  & $15{\substack{+14 \\ -0}}$ & $4.40{\substack{+0.18 \\ -0.17}}$ & $0.67{\substack{+0.07 \\ -0.0}}$ & 134.8 \\
     \\
     &  & 2 &  & $8.97\pm0.24$ & $1.01\pm0.03$ & $1.70\pm0.06$ & $0.31\pm0.03$ & 1.7 & $12{\substack{+32 \\ -0}}$ & $4.43{\substack{+0.16 \\ -0.07}}$ & $0.62{\substack{+0.14 \\ -0.0}}$ & 135 \\
     \\
     
     NGC6618-B0 & OI 345 & 1 & O6V\tablefootmark{a} & 7.4 &  &  &  &  & $44{\substack{+7 \\ -30}}$ & $4.42{\substack{+0.07 \\ -0.50}}$ & $0.76{\substack{+0.01 \\ -0.10}}$ & 71.8 \\
    \\
     &  & 2 &  & $10.89\pm0.63$ & $20.77\pm1.51$ & $34.88\pm2.54$ & $0.04\pm0.01$ & 1.6 & $33{\substack{+5 \\ -26}}$ & $4.49{\substack{+0.10 \\ -0.15}}$ & $0.74{\substack{+0.0 \\ -0.20}}$ & 72.0\\
     \\
     &  & 3 &  & $12.40\pm1.25$ & $6.58\pm2.09$ & $11.05\pm3.51$ & $0.010\pm0.005$ & 1.6 & $28{\substack{+23 \\ -10}}$ & $4.36{\substack{+0.2 \\ -0.1}}$ & $0.73{\substack{+0.04 \\ -0.05}}$ & 75.3 \\
    \\

     NGC6618-B260 & CEN 18 & 1 & O6V\tablefootmark{a} & 7.8 &  &  &  &  & $20{\substack{+32 \\ -5}}$ & $4.44{\substack{+0.04 \\ -0.29}}$ & $0.70{\substack{+0.07 \\ -0.03}}$ & 84.3 \\
     \\
     &  & 2 &  & $10.69\pm0.38$ & $32.55\pm2.45$ & $56.69\pm4.12$ & $0.07\pm0.01$ & 2.3 & $10{\substack{+27 \\ -1}}$ & $4.41{\substack{+0.2 \\ -0.2}}$ & $0.57{\substack{+0.2 \\ -0.01}}$ & 84.1\\
     \\
     &  & 3 &  & $12.05\pm0.63$ & $20.70\pm4.76$ & $34.76\pm7.99$ & $0.020\pm0.005$ & 2.3 & $20{\substack{+17 \\ -0.0}}$ & $4.51{\substack{+0.09 \\ -0.01}}$ & $0.68{\substack{+0.07 \\ -0.0}}$ & 85.6 \\
     \\

    \hline
    \end{tabular}
    \tablefoot{
\tablefoottext{a}{\citet{Hoffmeister+2008};}
\tablefoottext{b}{Combined magnitude of CEN~1a and 1b;}
\tablefoottext{c}{\citet{Ramirez+2017};}
\tablefoottext{d}{\citet{Broos+2007}.}}
    
\end{sidewaystable*}

%
%

\end{appendix}

\end{document}